\documentclass[aoas]{imsart}

\RequirePackage{amsthm,amsmath,amsfonts,amssymb}
\RequirePackage[authoryear]{natbib}
\RequirePackage{graphicx}
\RequirePackage{float}
\RequirePackage{subcaption}
\usepackage{algpseudocode}
\usepackage{algorithm}
\usepackage{bbm}

\usepackage{xcolor}

\startlocaldefs
\theoremstyle{plain}

\theoremstyle{remark}

\endlocaldefs

\begin{document}

\begin{frontmatter}
\title{Can Generalized extreme value model fit the real stocks}
\runtitle{GEV with large perturbation}

\begin{aug}
\author[A]{\fnms{Sen}~\snm{Lin}\ead[label=e1]{slin31@uh.edu}},
\author[B]{\fnms{Ao}~\snm{Kong}\ead[label=e2]{aokong@nufe.edu.cn}
}
\and
\author[A]{\fnms{Robert}~\snm{Azencott}\ead[label=e3]{robertazencott@gmail.com}}
\address[A]{Department of Mathematics,
University or Houston\printead[presep={,\ }]{e1,e3}}

\address[B]{Department of Finance,
Nanjing University of Finance and Economics\printead[presep={,\ }]{e2}}
\end{aug}

\begin{abstract}
The Generalized Extreme Value (GEV) distribution plays a critical role in risk assessment across various domains, such as hydrology, climate science, and finance. In this study, we investigate its application in analyzing intraday trading risks within the Chinese stock market, focusing on abrupt price movements influenced by unique trading regulations. To address limitations of traditional GEV parameter estimators, we leverage recently developed robust and asymptotically normal estimators, enabling accurate modeling of extreme intraday price fluctuations. We introduce two risk indicators: the mean risk level ($mEVI$) and a Stability Indicator (STI) to evaluate the stability of the shape parameter ($\xi$) over time. Using data from 261 Chinese and 32 U.S. stocks (2015–2017), we find that Chinese stocks exhibit higher $mEVI$, corresponding to greater tail risk, while maintaining high model stability. Additionally, we show that Value at Risk (VaR) estimates derived from our GEV models outperform traditional GP and normal-based VaR methods in terms of variance and portfolio optimization. These findings underscore the versatility and efficiency of GEV modeling for intraday risk management and portfolio strategies.

\end{abstract}


\end{frontmatter}

\section{Introduction}
The Generalized Extreme Value (GEV) distribution is widely used for risk assessment across multiple fields. In hydrology, it helps analyze rare rainfall or flood events, which is crucial for effective water resource management and infrastructure planning. In climate science, the GEV distribution aids in understanding extreme temperature fluctuations and storms, providing insights into the impacts of climate change. In finance, it models extreme fluctuations in asset prices and market returns, assisting in portfolio risk evaluation and management strategies.

Among GEV characteristics, the shape parameter $\xi$, also called the Extreme Value Index (EVI), and GEV extreme quantiles play crucial roles in evaluating risk levels \cite{nolde2021}. When $\xi > 0$, a larger $\xi$ indicates a heavier upper tail for the GEV distribution, thus signaling a greater out-of-sample positive risk. Conversely, for negative $\xi$, a smaller $\xi$ denotes a larger out-of-sample negative risk. However, many published estimators of the three GEV parameters have practical limitations. For instance, the maximum likelihood estimator (MLE) of $\xi$ is not asymptotically normal when $\xi < -0.5$, and the probability weighted moment estimator (PWM) does not exist when $\xi > 0.5$. These limitations complicate accurately fitting GEV distributions to observable extreme events in real-time series. Our companion paper (\cite{Lin2024} introduces a more widely usable class of estimators for fitting GEV distributions to time series.

In the context of the Chinese stock market, which is the world’s second-largest, unique regulations such as "T+1" trading and "price limits" are in place. "T+1" trading means that traders cannot sell shares bought on the same day, though they can repurchase previously sold shares. "Price limits" regulations halt trading for a stock if its price rises or falls by 10\% within a day, aiming to reduce volatility and protect individual investors. These rules may unintentionally cause significant overnight price changes. Although there is extensive research on overnight returns (\cite{guo2012}, \cite{wu2015}, \cite{qiao2020}), understanding intraday trading behavior remains an important question. Despite regulations, intraday trading still exhibits significant volatility, with notable price movements occurring throughout the day (\cite{kong2021}). Our study aims to comprehensively assess the distribution of abrupt and extreme price movements intraday using the robust GEV fitting algorithms developed in Lin (2024).

Our investigation revolves around two central questions: evaluating intraday trading risks in Chinese stocks and quantifying the stability of risk models over time. For a stock, we first use the block maxima (BM) method to extract extreme values in the standardized log returns observed across a moving time window and fit a GEV distribution with parameters $\hat\theta(t)=\left(\hat\xi(t),\hat\mu(t), \hat\sigma(t)\right)$ to the block maxima observed in the time window $W(t-k,t)$ of fixed duration $k$. This yields the estimation $EVI(t)$. The mean risk level for the stock is quantified by the mean value $mEVI$ of $EVI(t)$ over time.

Stability of risk models over time has also been an active field of research. In this paper, we mainly focus on the stability of a key GEV parameter, the shape parameter EVI. The stability of EVI over time has been studied, for instance, by \cite{quintos2001} and \cite{dehaan2021} for the 1997-1998 Asian financial crisis and for the SP500 loss, respectively. These two papers modeled extreme values using the peak-over-threshold method and used the CUMSUM statistic \cite{page1954} to detect changes in EVI. However, their sequences of EVI estimates are highly auto-correlated, violating the CUMSUM assumptions. To avoid these technical problems, we introduce and compute a Stability Indicator (STI) to quantify the stability of EVI estimates over time.

We computed these two risk indicators, $mEVI$ and STI, for each stock in two stock pools: 32 actively traded U.S. stocks and 261 Chinese stocks (Jan 2015 - Aug 2017). We find that these Chinese stocks generally exhibit higher {mean risk level $mEVI$}, which corresponds to thicker GEV right tail and hence to higher risk levels. Interestingly, a high proportion of these Chinese stocks exhibit fairly stable risk models over time, suggesting a high degree of risk predictability.

Value at Risk (VaR) is another commonly used method for quantifying risks in investment portfolios. Essentially, by focusing on the upper tail of the loss distribution, VaR answers the question: "What is the maximum loss that could be expected with a specified level of confidence over a set time horizon?". Extreme quantiles, such as the $99\%$ quantile, have been applied in various contexts for VaR analysis of financial positions. The normal distribution has often been applied to model stock returns and estimate the VaR (Normal-VaR). In extreme value theory approaches, the Generalized Pareto (GP) distribution was typically used to estimate the VaR (GP-VaR). Examples of this GP approach include exchange markets (\cite{danielsson1997}), stock market indices in developed markets (\cite{longin2000}) as well as emerging markets (\cite{gencay2004}), treasury yields (\cite{bali2003}), and electricity markets (\cite{chan2006}). Our findings indicate that fitting GEV distributions to intraday stock returns by our asymptotically normal estimators of GEV parameters, does also provide efficient VaR estimates (GEV-VaR); indeed, via extensive simulations on multiple Heston joint SDEs, we show that GEV-VaR estimates have smaller variance than GP-VaR estimates (for the same number of observations). This led us to develop a simplified VaR-based strategy to optimize portfolio in the Chinese and U.S. stock market, to concretely compare efficiencies between different VaR estimates: our GEV-VaR estimates reached better mid-term portfolio valuations than Normal-VaR estimates. We aim to highlight the versatility of the GEV approach in this context.

The structure of our paper is as follows: First, we introduce the research framework, including estimation and evaluation methods for constant EVI, in Section \ref{section: framework}. In Section \ref{section: real data}, we apply this pipeline to analyze EVI and VaR across 261 stocks in the Chinese market from January 5, 2015, to August 31, 2017, and compare these results with 32 U.S. stocks from the same period. A simple portfolio optimization strategy based on VaR is also proposed to highlight the versatility of the GEV distribution. We then validate our analytical pipeline using stationary simulated Heston SDE trajectories. Finally, we discuss detecting large jumps through the fitted GEV model, providing insights into intraday trading risk dynamics.

\section{Framework for extreme returns analysis in stock market}\label{section: framework}
\subsection{Multi-Quantile estimator for GEV parameters}\label{section: multi-q}
Consider a sequence of i.i.d random variables $Y_1,...,Y_n$ having the same GEV distribution, with CDF $G_{\theta}$ given by the classical formula :
\begin{equation*}
G_{\theta}(y) =
\begin{cases}
\exp\left(-\left(1 + \xi\left(\frac{y - \mu}{\sigma}\right)\right)^{-1/\xi}\right), & \xi \neq 0 \\
\exp\left(-\exp\left(-\frac{y - \mu}{\sigma}\right)\right), & \xi = 0
\end{cases}
\end{equation*}

And the $q$th-quantile $Q$ of $G_\theta$ is given by:
\begin{align}\label{eq:quantile}
Q=\begin{cases}
    \mu + \frac{\sigma}{\xi} \left(\exp(-\xi LL)-1\right) &\text{when}\ \xi\neq 0 \\
    \mu - \sigma LL &\text{when}\ \xi = 0
\end{cases}
\end{align}
where $LL=\log(-\log(q))$.

Select any triplet of percentiles $0<q_1<q_2<q_3<1$, denote $Q_1,Q_2,Q_3$ the corresponded quantile of the distribution $G_\theta(y)$ and $\hat Q_1,\hat Q_2,\hat Q_3$ the associated empirical quantile of $Y_1,...Y_n$. A Three-Quantile estimator of $\xi$ is introduced by \cite{Lin2024}. Multi-Quantile estimators were then developed to address the low estimation accuracy problem in Three-Quantile estimators in the same literature. Consider $p$ Three-Quantile estimators of $\xi$ defined by $p$ distinct triples of quantiles and denoted as $\hat\xi_1,...\hat\xi_{p}$. We construct our optimized weighted sum of these $p$ estimators as a final estimate of $\xi$.
\begin{equation}
    \hat\xi = \sum_j^p w_j\hat\xi_j
\end{equation}
The weights $w_j$ are obtained from optimizing the theoretical estimation error of $\xi$. For detailed computation refer to \cite{Lin2024}. The estimation error of the weighted sum is explicitly computable and asymptotically normal for any $\xi\in\mathbb R$. This is the key point in quantifying the stability of risk model in Section \ref{section: STI}. The estimates $\mu$ and $\sigma$ can also be determined and their estimation errors are asymptotically normal as well as explicitly computable.

\subsection{Block maxima method}\label{section: BM}
Let $Z_1,Z_2...,$ be an i.i.d sequence of random variables with continuous distribution function $F$. For $m=1,2,...$ and $i=1,2,...,n$, define the block maxima
\begin{equation}
    {Y_i} = \max_{(i-1)m<s\leq im}Z_s
\end{equation}
The whole sequence of $N =mn$ observations are divided into $n$ blocks of size $m$. Assume $F$ belongs to the max-domain of attraction $D(G_\xi)$ of a GEV distribution with EVI $\xi$. 

When applying the Multi-Quantile estimator of $\xi$ on the sequence of block maxima $Y_i$, we then have the following asymptotic normality
\begin{equation}\label{eq: BM}
    \sqrt{n}(\hat\xi-\xi) \sim \mathcal{N}\left(B(\xi, m),\Sigma_\xi\right)
\end{equation}
as $m\to\infty$, $n\to\infty$ and $N\to\infty$.
Where the bias term $B(\xi,m)$ is a function of $\xi$ and $n$. And $\sqrt{n}B(\xi,m)\to\lambda(\xi)\in\mathbb R$. Though the bias term $B(\xi,m)$ is generally hard to compute, the $\Sigma_\xi$ can be calculated explicitly (\cite{Lin2024}). 
The computation of $\Sigma_\xi$ is crucial in quantifying the stability of risk model. We elaborate this in the section \ref{section: STI}.


\subsection{Goodness of fit}
We use two simultaneous criteria to investigate GEV distribution correctly models the observed sequence of block maxima. The first criterion is based on the \emph{p}-value of the two-sample KS test to quantify the goodness of fit of GEV distribution to the observed block maxima. We systematically require that KS \emph{p}-value $> 0.05$ to assert that with more than $95\%$ confidence these block maxima can be fitted by the GEV distribution with estimated parameters $\hat\xi, \hat\mu, \hat\sigma$. 

The second criterion reflects that we are fitting a GEV model to positive random variables in this paper. So we will also systematically compute our Model Positivity Index (MPI) defined by $MPI = P(Y<0) = G_{\hat\theta}(0)$. We will require that $MPI < 10^{-4}$ to validate an estimated GEV model.

\subsection{Empirical check whether $\xi(t)$ is constant over time}\label{section: STI}
Consider the sequence of extracted block maxima $\{Y_t\}_{t=1,2,...,T}$ as a time series such that $Y_t$ has a GEV distribution $G_{\xi(t),\mu(t),\sigma(t)}$. The $\xi(t)$ is estimated locally from observations in the moving time window $W(t-k,t)$, which gathers the $k$ observations $Y_t,Y_{t-1},...,Y_{t-k}$. {In our risk analysis on intraday data for two pools of stocks, we extract 10-min log-returns maxima every 2 days, {then the block size} $m=46$ for the Chinese stock market and $m=76$ for the U.S. stock market; the moving window length $k$ is roughly 1 trading year.}

It is of practical interest to test whether the extreme value index $EVI(t) = \hat\xi(t)$ is statistically constant over time, since this will enable better predictions of extreme events and responses to their occurences. If the "true" $\xi(t)$ is equal to some fixed $\xi_0$ for all $t$, then the estimate $\hat\xi(t)$ should be statistically constant. Under the assumption $\xi(t) = \xi_0$ for all $t=1,2,...,T$, we expect to have, 
\begin{equation*}
    \sqrt{k}(\hat\xi(t)-\xi_0) \sim \mathcal{N}\left(B(\xi_0, m),\Sigma(\xi_0)\right)
\end{equation*}
Different from the equation \ref{eq: BM}, the bias term $B(\xi_0, m)$ is a function of $\xi_0$ and $m$ and the $\Sigma(\xi_0)$ is a explicit function of $\xi_0$.

We test the constancy hypothesis $\xi(t) = \xi_0$ by setting $\xi_0$ to be the empirical mean $\bar\xi$ of all the $\hat\xi(t)$. At time $t$, we consider $\hat{\xi}(t)$ to be statistically equal to $\bar\xi$ if $|\hat{\xi}(t) - \bar\xi| \leq \text{error margin of}\ \bar\xi$. The asymptotic distribution of $\hat\xi(t)$ is crucial for evaluating this error margin. We compute the error margin using explicit formulas derived for the asymptotic normality parameters of our Multi-Quantiles estimator of $\xi\in\mathbb{R}$ \cite{Lin2024}. These asymptotic formulas make the constancy test feasible. The error margin of $\bar\xi$ is computed at a $95\%$ confidence level.

We define a \emph{stability indicator} ($STI$) to quantify the stability of $EVI(t) = \hat{\xi}(t)$. For each stock, $STI$ is the observed frequency of the time points $t$ such that $|\hat{\xi}(t) - \bar\xi|$ is less than the error margin of the time average $\bar\xi = mEVI$. A higher $STI$ indicates greater stability for the risk model. For any stock with $STI > 80\%$, we will say that $EVI(t)$ is roughly constant over time and that our GEV risk model has good stability over time (see Section 4 for practical comments on requiring $STI > 80\%$).

\subsection{Value at Risk}
In financial risk management, understanding and quantifying portfolio risk is crucial. One commonly used measure is Value-at-Risk (VaR), a statistical technique that estimates the potential loss in value of a portfolio over a defined period for a given confidence interval. Essentially, VaR answers the question: "What is the maximum loss that could be expected with a specified level of confidence (e.g., 1\% confidence level) over a set time horizon?" By focusing on the upper tails of the loss distribution, VaR is estimated using a high quantile, such as the 99\% quantile. In early publications and frequent practice, a normal distribution is often fitted to the observed returns of a given stock, and the so-called Normal-VaR of the stock is estimated by, for example, the 99\% quantile of the fitted normal distribution in \cite{jpmorgan1996, duffie1997}. Additionally, the upper tail of the loss distribution has often been modeled by a GP distribution (GP-VaR) in \cite{nolde2021}.

In this article, we focus on the tail behavior of absolute returns after natural short-term standardization. By concentrating on absolute returns, we naturally model extreme potential losses as well as extreme potential gains. This approach provides a more comprehensive view of the occurrence of large instantaneous volatility for a given stock. Over the time window $W(t-k,t)$, as mentioned in Section \ref{section: STI}, we compute our estimate GEV-VaR of the Value at Risk as the 99\% quantile of the fitted GEV distribution $G_{\hat\theta(t)}$. For window $W(t-k,t)$, this yields an estimate $VaR(t)$, given by the explicit formula:
\begin{equation}\label{eq: var}
    VaR(t) = \hat\mu(t) + \frac{\hat\sigma(t)}{\hat\xi(t)} \left(\exp(-\hat\xi(t) \log(-\log(0.99)))-1\right)
\end{equation}
All the numerical $VaR(t)$ mentioned later in this paper are evaluated as above by our GEV-VaR approach.

\section{Analysis of extreme values for stock returns}\label{section: real data}
In this section, we selected 261 Chinese stocks and 32 U.S. stocks, which were actively traded from January 2014 to August 2017, to construct separate Chinese and U.S. stock pools. Within each pool, we analyzed the extreme values of each individual stock using the computational pipeline outlined in Section \ref{section: framework} to concretely illustrate our approach. We then studied the time evolution of the shape parameter $EVI(t)$ for our dynamically fitted GEV models, as well as our GEV-VaR estimates $VaR(t)$, on both individual stocks and on a pool-wide basis. This analysis showcases the different dynamics of extreme value models across the two stock pools, providing a prototype that can be extended to the entire market for studying dynamic structural changes. Finally, we used the candidate stocks in each pool to construct Chinese and U.S. portfolios, proposing a weight calculation method based on VaR. We demonstrated the advantage of the GEV-VaR over the Normal-VaR in portfolio optimization.

\subsection{Financial data set and data processing}\label{section: processing}
Our study is based on the level-2 transaction data of Chinese stock market and U.S. stock market. The level-2 data, updated every 3 seconds, encompass the best ten quotes, current transaction price and volume, as well as cumulative trade and volume information from the preceding record.

\textbf{Chinese stock market} Main-board and second-board stocks on the Shenzhen Stock Exchange in China, covering the period from January 2014 to August 2017 (a total of 896 trading days, 244 trading days in one year). These data were sourced from the Wind database. A comprehensive dataset of 2019 stocks was assembled for analysis. The Shenzhen Stock Exchange operates from 9:30 am to 11:30 am and 1:00 pm to 3:00 pm, comprising a total of 4 hours per trading day. 

\textbf{U.S. stock market} 30 component stocks of the Dow Jones Industrials Average and 12 active stocks previously included in the Dow Jones Industrials Average are included (see detailed stock tickers in Appendix), covering the period from January 2014 to August 2017 (a total of 924 trading days, 252 trading days in one year). We only consider data collected during the regular US trading session, which start at the 9:30 a.m. and ends at the 4:00 p.m., with 6.5 hours per trading day.

For both data set, only minutes with non-zero trading volume are included. Bars with zero volume are excluded. And we only focus on the 10-minute frequency level prices. Sessions exhibiting moderate or high trading activity are exclusively considered. To ensure data integrity, days featuring more than 3 consecutive missing prices were omitted from our analysis. Similarly, days with over 30 consecutive minutes devoid of recorded price movement, as well as those with less than 90 minutes of recorded price movement (i.e., returns not equal to 0 or missing), were excluded. After filtering, only 261 stocks in Chinese market and 32 stocks U.S. market with active trading occurring on over $90\%$ of the total trading days are included in the study.

\subsection{Standardized return}
High-frequency returns are known to have approximately zero mean but a strongly fluctuating variance, with both intraday seasonalities and long-memory, intermittent dynamics mention by \cite{boudt2014intraday}. Assume stock prices $P_{h_i}$ are observed at time $h_i$, $i= 1,2,3,...$, where $h_i-h_{i-1}=10$ minutes. Define the log-return $LR_{h_i} = \log(P_{h_{i+1}})-\log(P_{h_i})$. We then define the standardized return $SLR_h$ as:
\begin{equation}
    SLR_{h_i} = \frac{LR_{h_i}}{std(LR_{h_i})f_{h_i}}
\end{equation}
where $std(LR_{h_i})$ is the standard deviation of return $LR_{h_i}$. It is estimated by
\begin{equation*}
    std(LR_{h_i}) = \sqrt{\frac{\pi}{2K}\sum_j^K|LR_{h_{i-j}}||LR_{h_{i-j+1}}|}
\end{equation*}(\cite{barndorff2004}). 
Note that we drop all overnight returns. Here $f_{h_i}$ is an estimator of the periodicity component (\cite{boudt2014intraday}) to account for the intraweek periodicity in high-frequency returns. We define the decorrelation time as the the time delay needed for the autocorrelation of $|SLR_{h_i}|$ to fall within the $95\%$ confidence interval $[-1.96 \times \frac{1}{\sqrt{n}},1.96 \times \frac{1}{\sqrt{n}}]$, where $n$ is the sample size. In the $261$ Chinese stocks, the decorrelation time for the $|SLR_{h_i}|$ ranges from $100$ to $550$ minutes. While in our U.S. $32$ stocks, the decorrelations takes $420$ to $600$ minutes. For safety, we consider the $|SLR_h|$ generally decorrelates in two days. 

\begin{figure}[!h]
  \centering
  \begin{subfigure}[b]{0.48\textwidth}
    \centering
    \includegraphics[width=\textwidth]{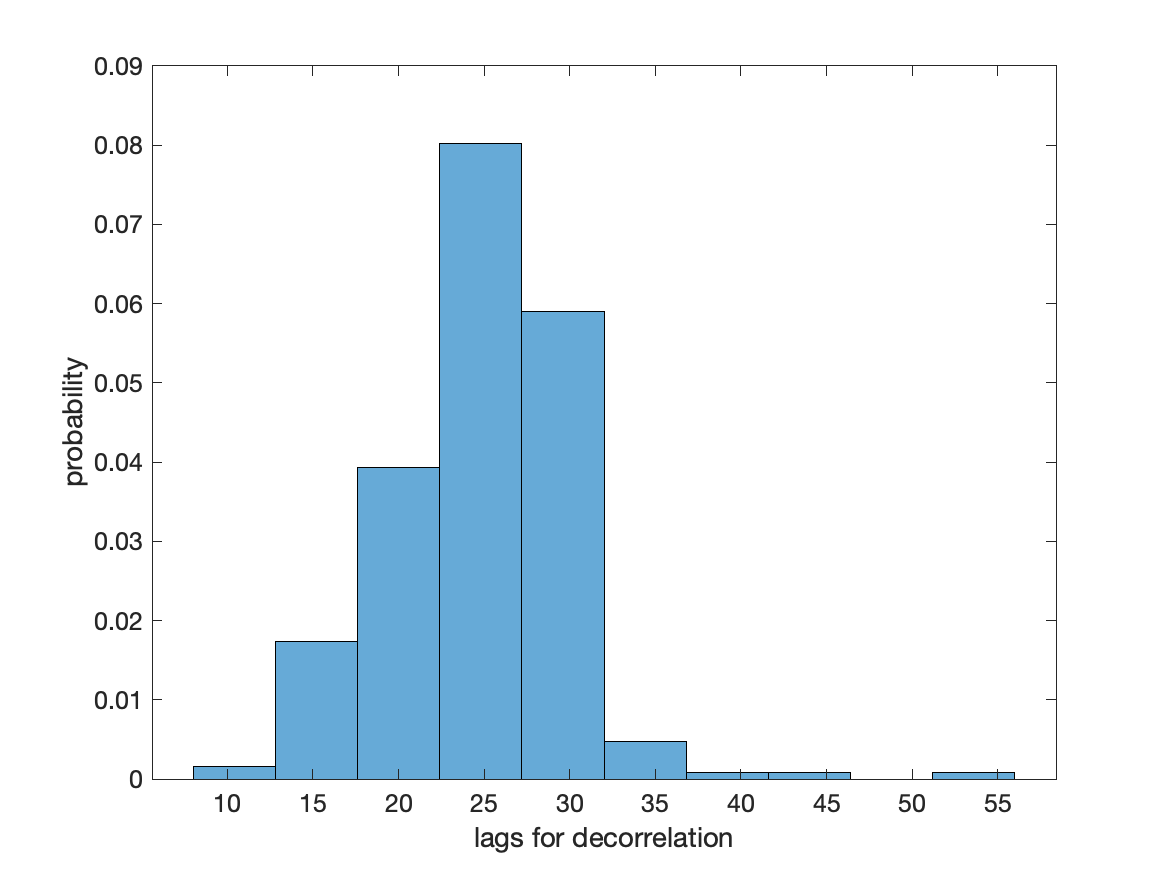}
    \caption{Chinese}
    \label{fig: chn_decorr}
  \end{subfigure}
  \hfill
  \begin{subfigure}[b]{0.48\textwidth}
    \centering
    \includegraphics[width=\textwidth]{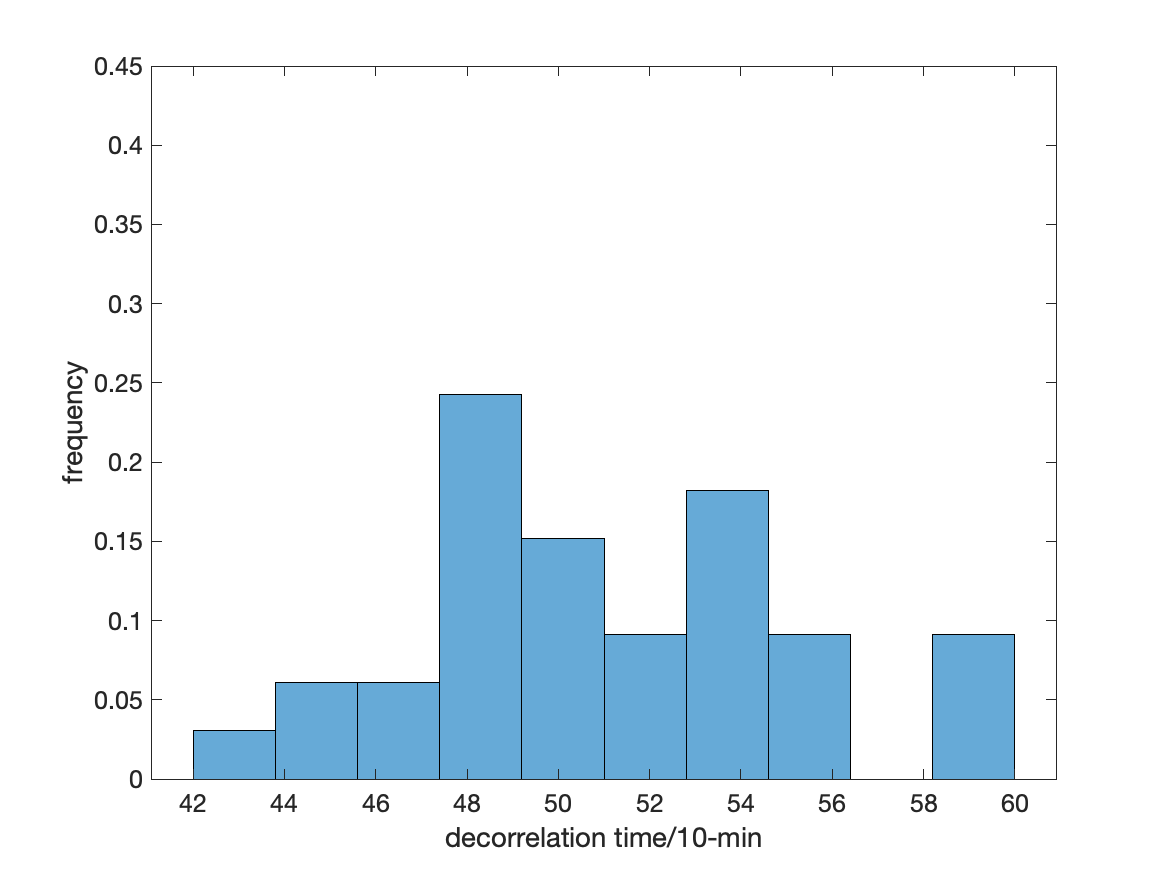}
    \caption{U.S.}
    \label{fig: us_decorr}
  \end{subfigure}
    \caption{Frequency of decorrelation time for portfolio of Chinese stocks and U.S. stocks. In the portfolio of $261$ Chinese stocks, the decorrelation time for the $|SLR_{h_i}|$ ranges from $100$ to $550$ minutes. While in the U.S. portfolio of $32$ stocks, the decorrelations takes $420$ to $600$ minutes.}
    \label{fig: decorrelation}
\end{figure}

\subsection{Dynamic of extreme value index for stocks in Chinese market}

We conduct the extreme value analysis in an online mode to mimic market monitoring. The trajectory of $|SLR_{h_i}|$, $i=1,2,3,...$ is naturally partitioned by trading days, $23$ observations per trading day. The stocks can be traded on day $t = 1,2,...,896 = T$. We compute the trading risk model by fitting a GEV model over the past one trading year (52 weeks). Specifically, we fit a GEV model based on a rolling window $W(t-k, t)=[t-245,t-244,...,t]$. To remove the impact of serial dependence in $SLR_{h_i}$, which will impact on the model fitting, we focus on the maximum $|SLR_{h_i}|$ of every two days. Recall that the auto-correlation function of $|SLR_h|$ diminishes in $1$ to $2$ days for stocks in Chinese pool (see Figure \ref{fig: decorrelation}). Each maxima is extracted from $46$ observations, and $123$ maximas will be used to fit a GEV model at time $t$. The monitoring starts from 1/5/2015 ($t=246$) to 8/31/2017 ($t=896$). We monitor the risk model every two days. For example, on 1/5/2015, we fit a GEV model by $123$ maxima $|SLR_h|$ extracted every two days from 1/2/2014 to 1/5/2015. Then, we monitor the risk model on 1/7/2015.

\begin{figure}[!h]
  \centering
  \begin{subfigure}[b]{0.48\textwidth}
    \centering
    \includegraphics[width=\textwidth]{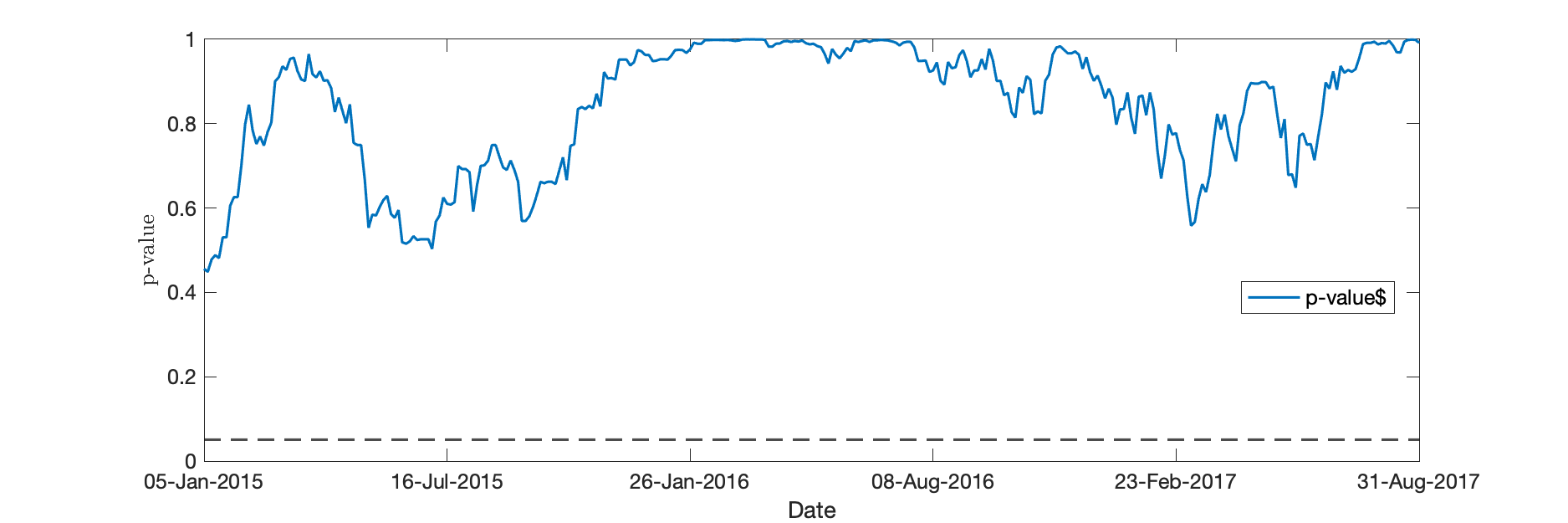}
    \caption{\emph{p}-value, Yanghe Brewy}
    \label{fig: yb_p}
  \end{subfigure}
  \hfill
  \begin{subfigure}[b]{0.48\textwidth}
    \centering
    \includegraphics[width=\textwidth]{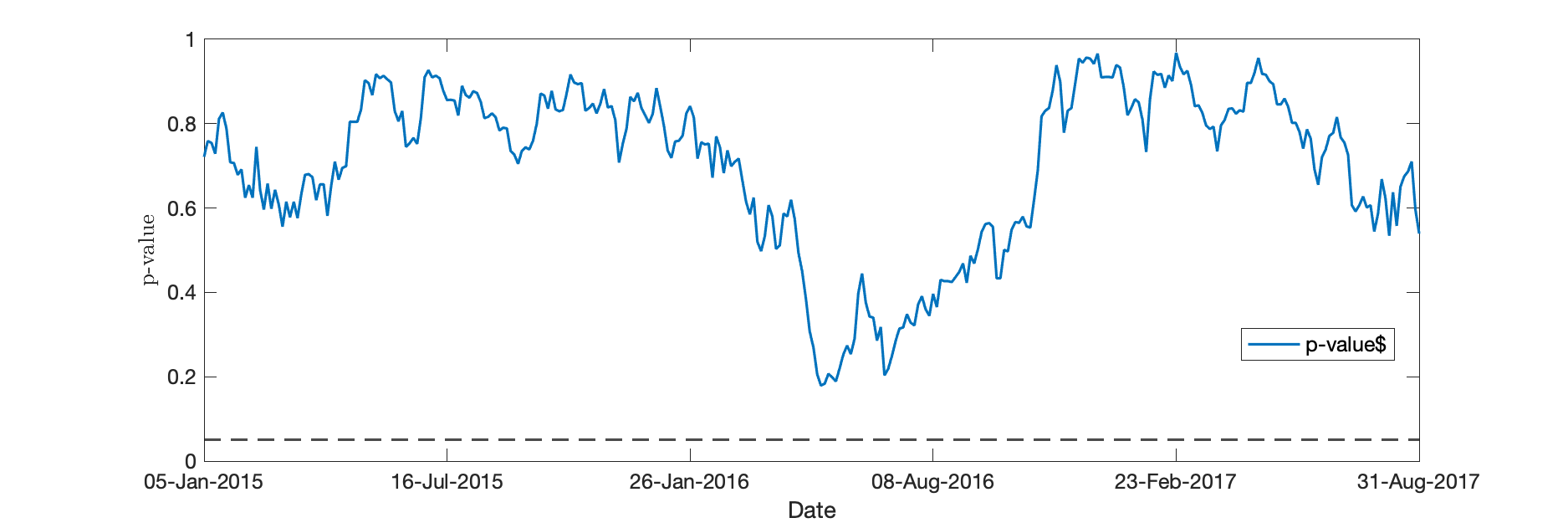}
    \caption{\emph{p}-value, Sino Wealth Electronic}
    \label{fig: swe_p}
  \end{subfigure}
  
  \begin{subfigure}[b]{0.48\textwidth}
    \centering
    \includegraphics[width=\textwidth]{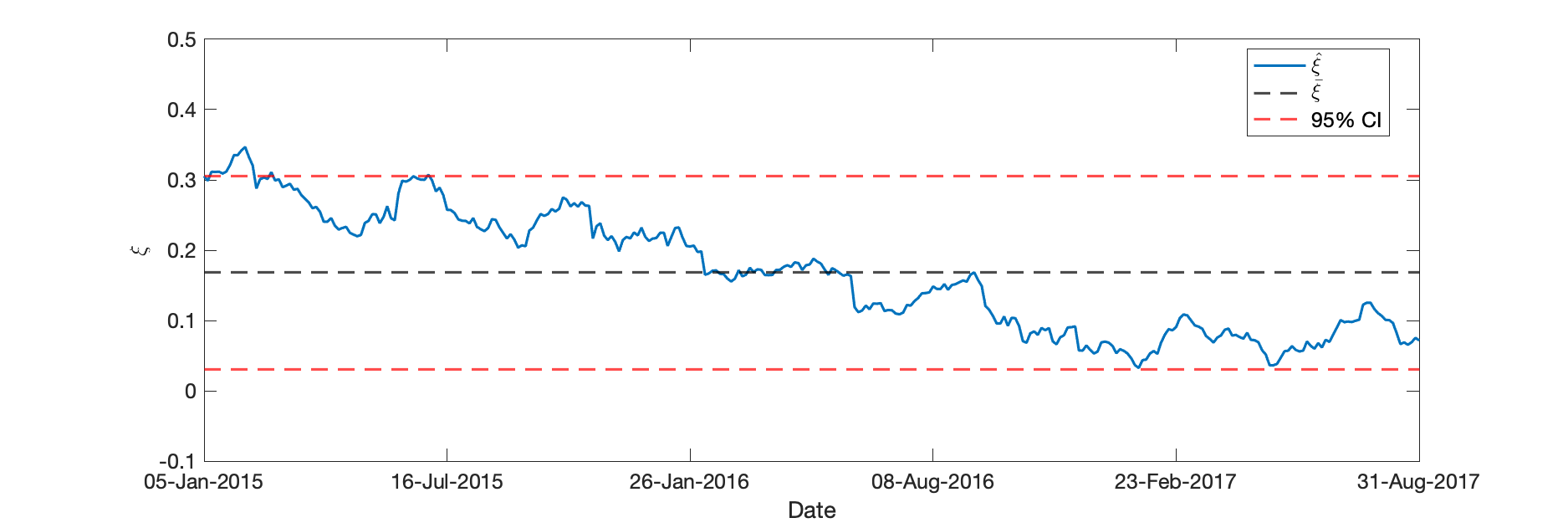}
    \caption{$EVI(t)$, Yanghe Brewy}
    \label{fig: yb_xi}
  \end{subfigure}
  \hfill
  \begin{subfigure}[b]{0.48\textwidth}
    \centering
    \includegraphics[width=\textwidth]{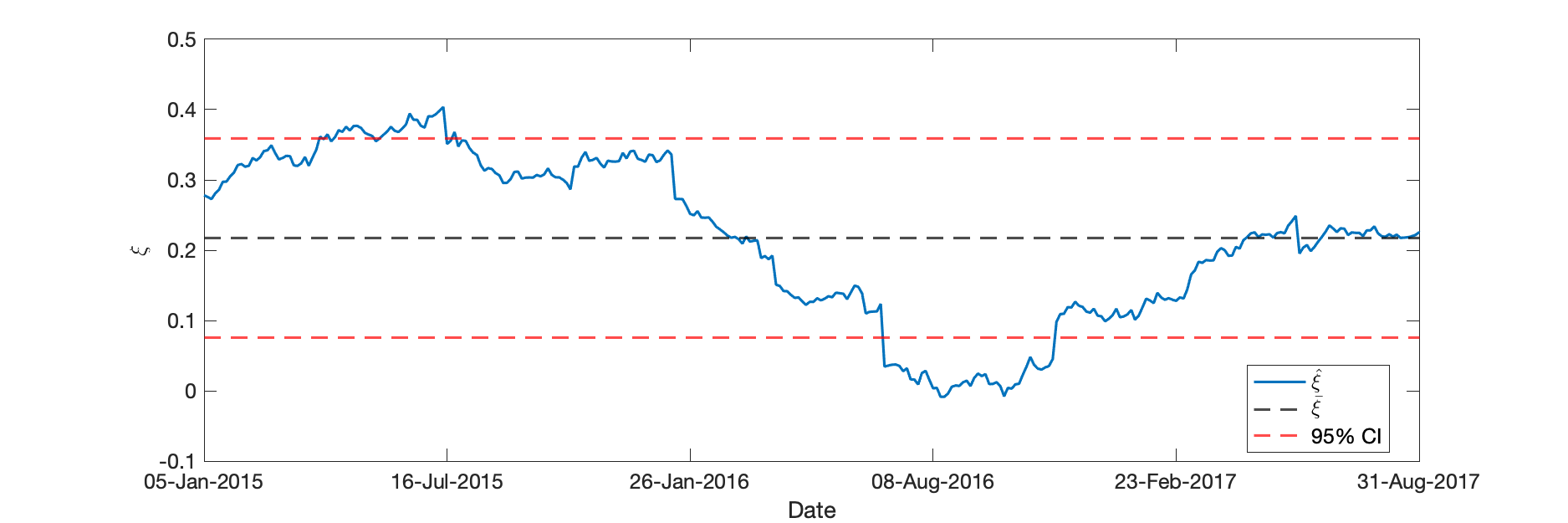}
    \caption{$EVI(t)$, Sino Wealth Electronic}
    \label{fig: swe_xi}
  \end{subfigure}

  \begin{subfigure}[b]{0.48\textwidth}
    \centering
    \includegraphics[width=\textwidth]{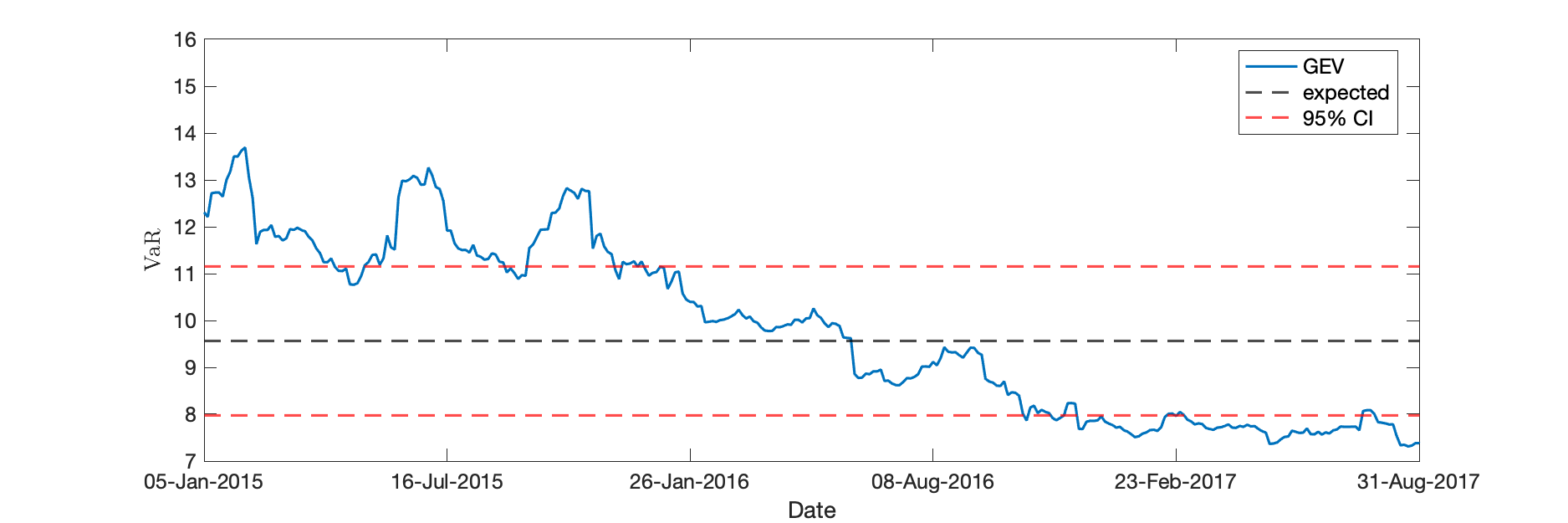}
    \caption{$VaR(t)$, Yanghe Brewy}
    \label{fig: yb_q99}
  \end{subfigure}
  \hfill
  \begin{subfigure}[b]{0.48\textwidth}
    \centering
    \includegraphics[width=\textwidth]{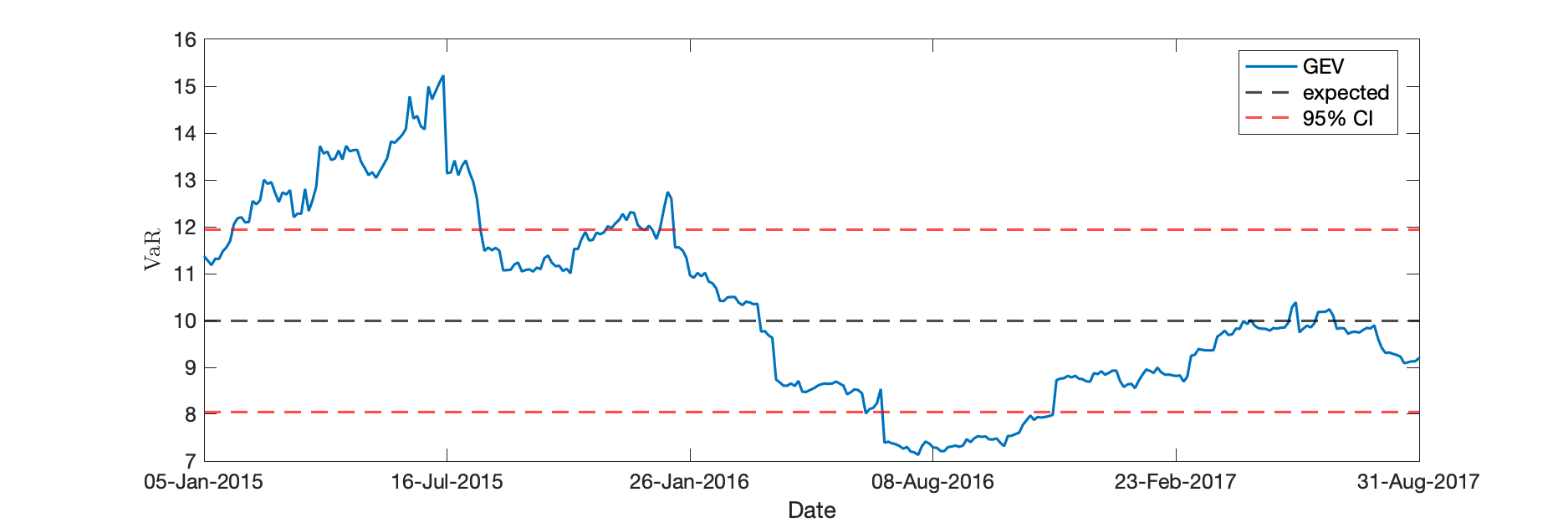}
    \caption{$VaR(t)$, Sino Wealth Electronic}
    \label{fig: swe_q99}
  \end{subfigure}
  \caption{Dynamic of \emph{p}-value, $EVI(t)$ and $VaR(t)$ for Sino Wealth Electronic and Yanghe Brewy stock from 1/5/2015 to 8/31/2017. The \emph{p}-value of KS test always shows good fit. The $EVI(t)$ fluctuate around its mean (black). And the $VaR(t)$ are high in 2015, a spike in $VaR(t)$ is observed during the May 2015 stock market crash in both stocks.}
  \label{fig: dynamic2}
\end{figure}

We randomly select stock 002304 Yanghe Brewy and stock 300327 Sino Wealth Electronic for detailed elaboration. For Yanghe Brewy stock, the GEV distribution can always model the maximum $SLR_h$ well (see, figure \ref{fig: dynamic2}), since the \emph{p}-value are always $>0.05$. In the mean time, $\sup_t MPI(t) = \sup_t G_{\hat\theta(t)}(0)=10^{-10}<<10^{-5}$. We thus conclude that the GEV distribution can always model the tail distribution of $|SLR_h|$ well. A strong decreasing trend is observed in $EVI(t)$, which generally indicates a less risky trading environment, but $EVI(t)$ remains statistically a constant as indicated by the empirical check ($STI = 95\%$), the $mEVI=0.16$. While the $VaR(t)$ is very high in 2015, since the market surged in 2014 and crashed in May 2015. A spike in $VaR(t)$ is observed during the stock market crash.

While for the Sino Wealth Electronic stock, only $78\%$ amount of time that equality between $EVI(t)$ and $mEVI=0.21$ is observed. There is at least one change in the risk model. One may use the change point detection algorithm to determine the time and frequency such change happens. We elaborate on how these changes can be detected in the section \ref{section: jump}. Besides, similar to the $VaR(t)$ of Yanghe Brewy stock, the $VaR(t)$ of Sino Wealth Electronic stock is also constantly high in and we observe a spike in $VaR(t)$ during the May 2015 Chinese stock market crash.

\subsection{Dynamic of extreme value index in stocks in U.S. market}

The extreme value analysis in U.S. stock market is conducted exactly as in Chinese stock market. Since the number of trading days in the U.S. stock market is slightly different from the Chinese market. Here $t = 1,2,...,924$. We still use the $|SLR_h|$ in past one trading year to fit a GEV model. But the rolling window $W(t-k,t)$ has a different size with day $t-251,t-250,...,t$. Based on the decorrelation time in Figure \ref{fig: decorrelation} for most stocks in U.S market, we compute the maxima of $|SLR_{h_i}|$ over successive disjoint intervals of 2 days each, in order to to make sure that these successive maxima are decorrelated. One may partition each window $W(t-k,t)$ into $151$ non-overlapping segments of size $2$. Each maximum is computed over the 76 prices recorded within two trading days and $151$ maximas will be used to fit a GEV model at time $t$. The risk model is monitored every two days as well. The monitoring starts from 1/3/2015 ($t=253$) to 8/31/2017 ($t=924$).

We select APPL and CVX for detailed analysis. For both APPL and CVX, the GEV distribution can always model the maximum $|SLR_h|$ well. And we found that $EVI(t)$ fluctuates around the $mEVI$ and within its $95\%$ confidence interval for both stocks. We can thus conclude that for both stocks, $EVI(t)$ is always statistically equal to a fixed $mEVI$. We have $mEVI = 0.31$ for APPL and $mEVI = 0.17$ for CVX. They both have high $VaR(t)$ before mid of 2016, which decrease after Jan 2017. The spike in $VaR(t)$ in hasn't been observed during May 2015 Chinese stock market crash. But the $VaR(t)$ increased around June 2015, since the impact of stock crash in Chinese stock market spread globally.

\begin{figure}[!h]
  \centering
  \begin{subfigure}[b]{0.48\textwidth}
    \centering
    \includegraphics[width=\textwidth]{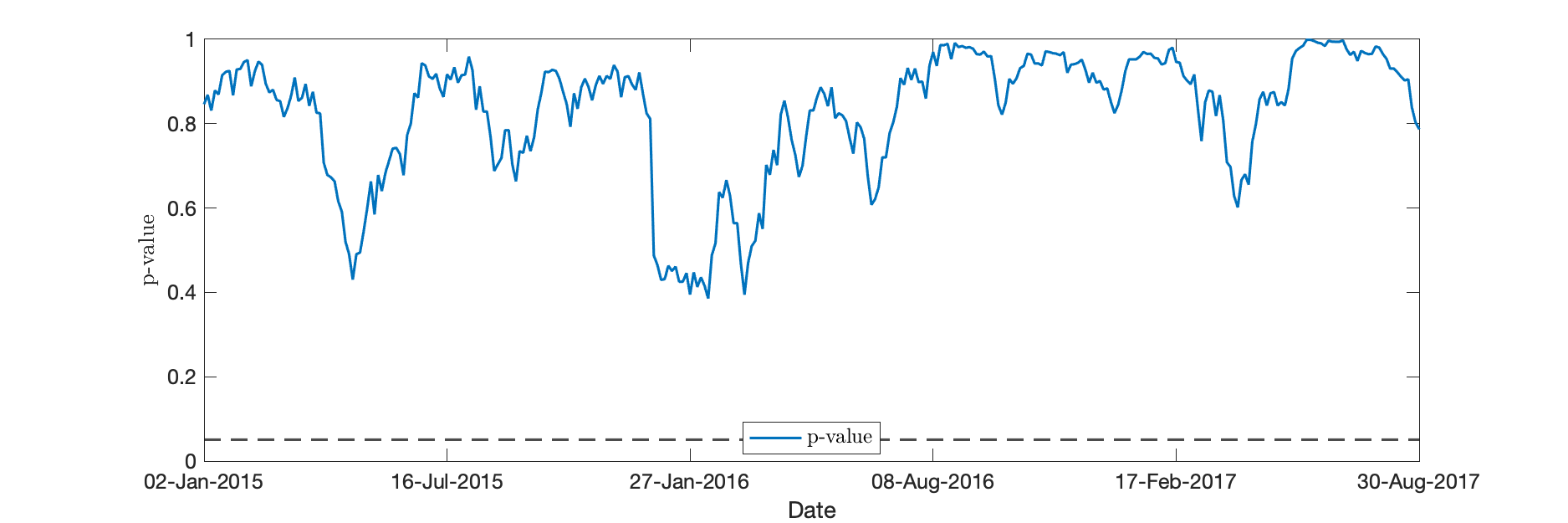}
    \caption{\emph{p}-value, APPL}
    \label{fig: appl_p}
  \end{subfigure}
  \hfill
  \begin{subfigure}[b]{0.48\textwidth}
    \centering
    \includegraphics[width=\textwidth]{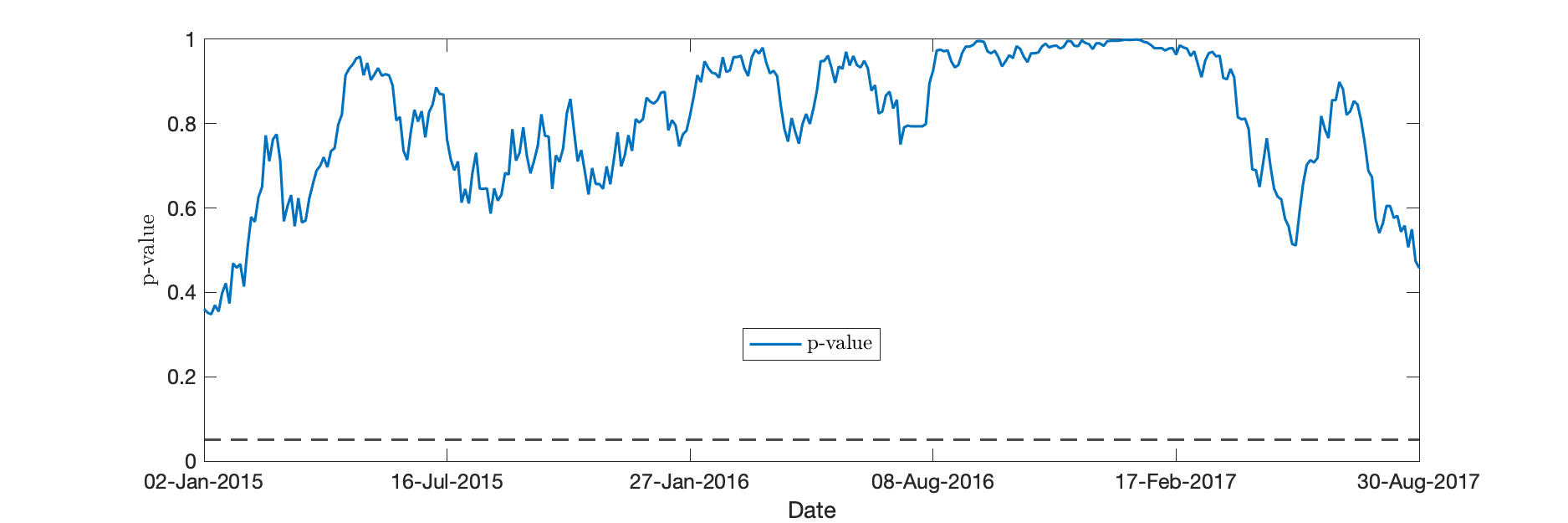}
    \caption{\emph{p}-value, CVX}
    \label{fig: cvx_p}
  \end{subfigure}
  
  \begin{subfigure}[b]{0.48\textwidth}
    \centering
    \includegraphics[width=\textwidth]{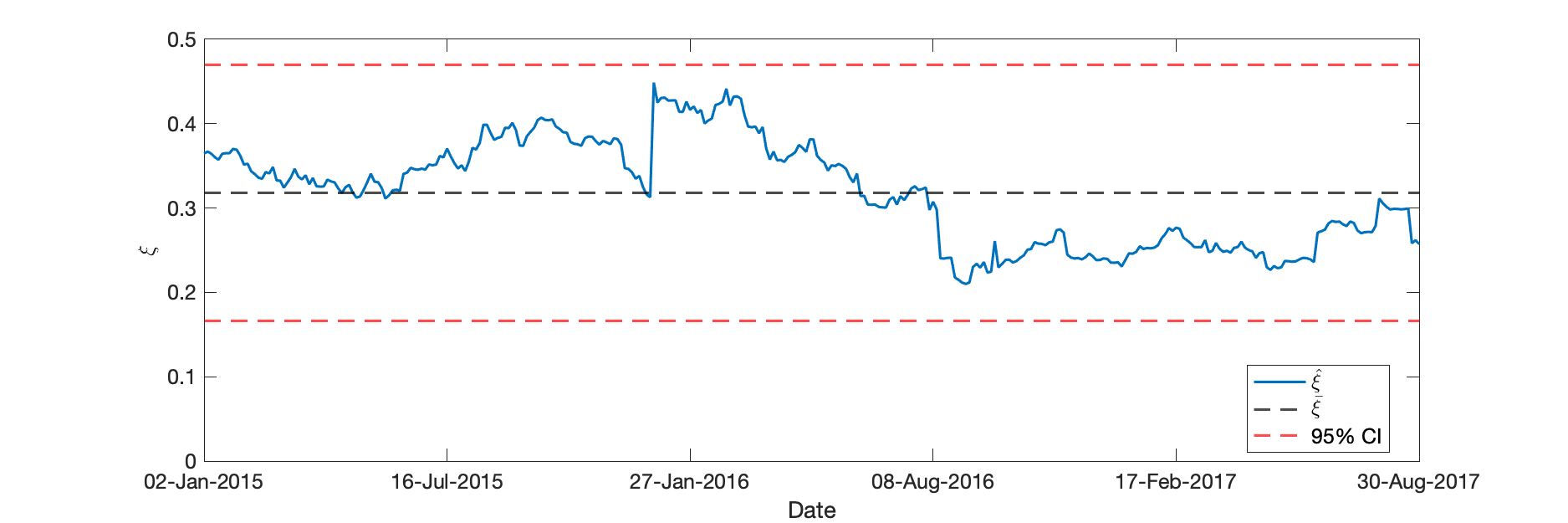}
    \caption{$EVI(t)$, APPL}
    \label{fig: appl_xi}
  \end{subfigure}
  \hfill
  \begin{subfigure}[b]{0.48\textwidth}
    \centering
    \includegraphics[width=\textwidth]{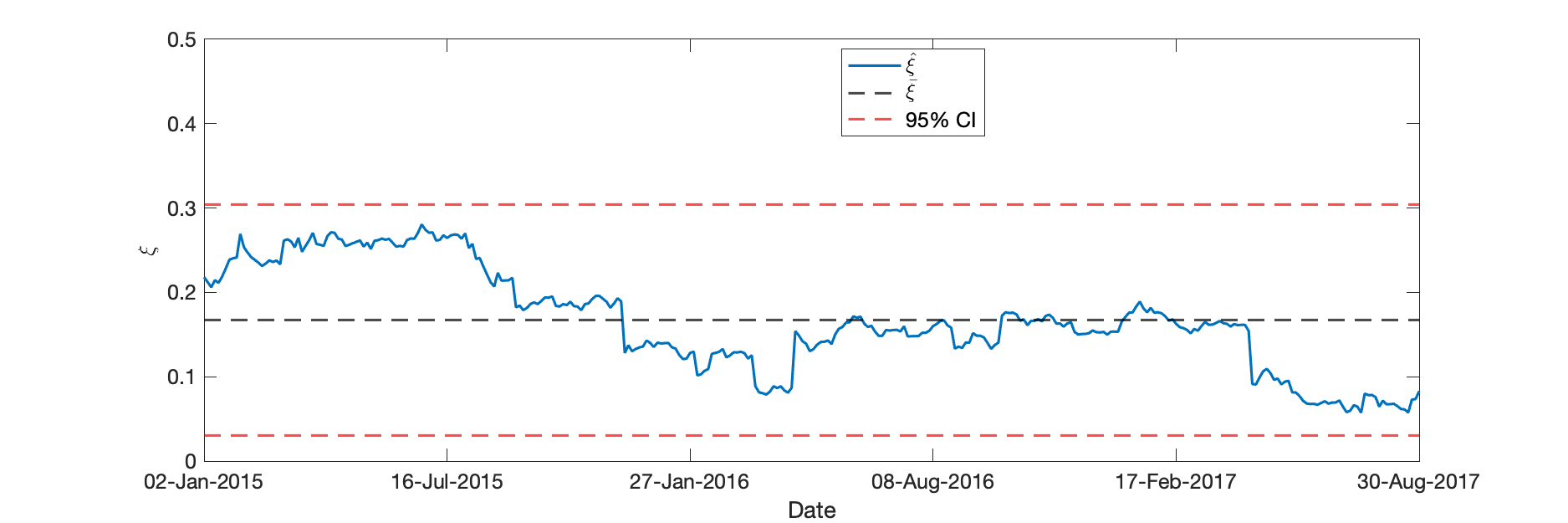}
    \caption{$EVI(t)$, CVX}
    \label{fig: cvx_xi}
  \end{subfigure}

  \begin{subfigure}[b]{0.48\textwidth}
    \centering
    \includegraphics[width=\textwidth]{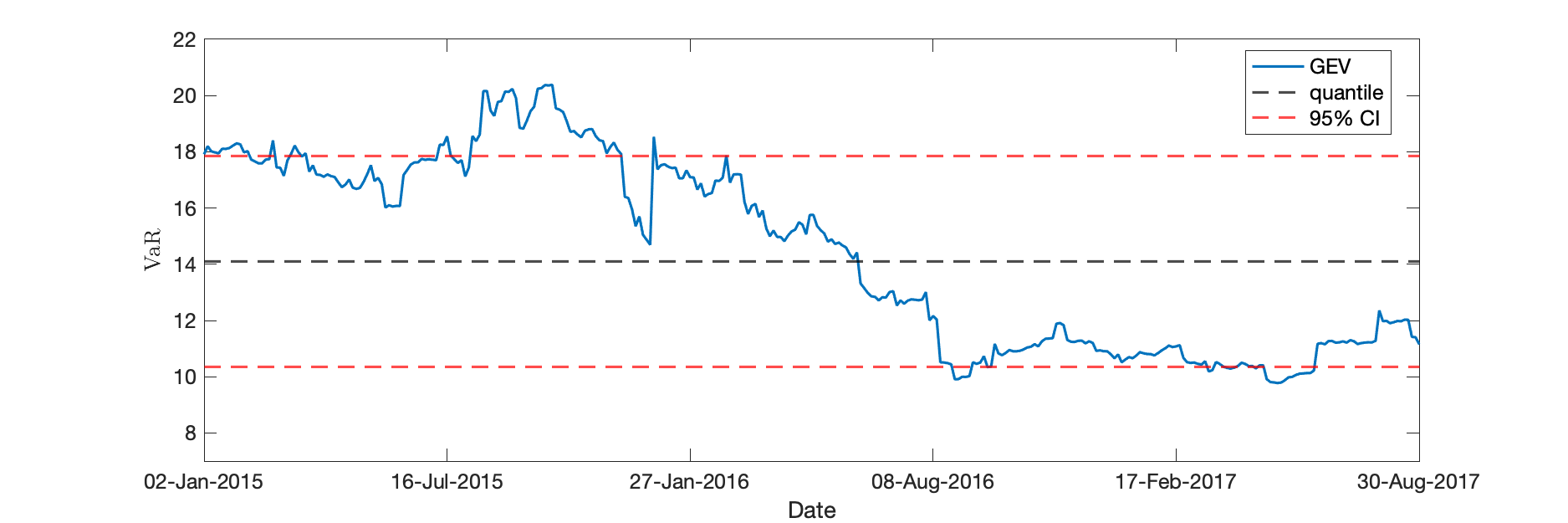}
    \caption{$VaR(s)$, APPL}
    \label{fig: appl_q99}
  \end{subfigure}
  \hfill
  \begin{subfigure}[b]{0.48\textwidth}
    \centering
    \includegraphics[width=\textwidth]{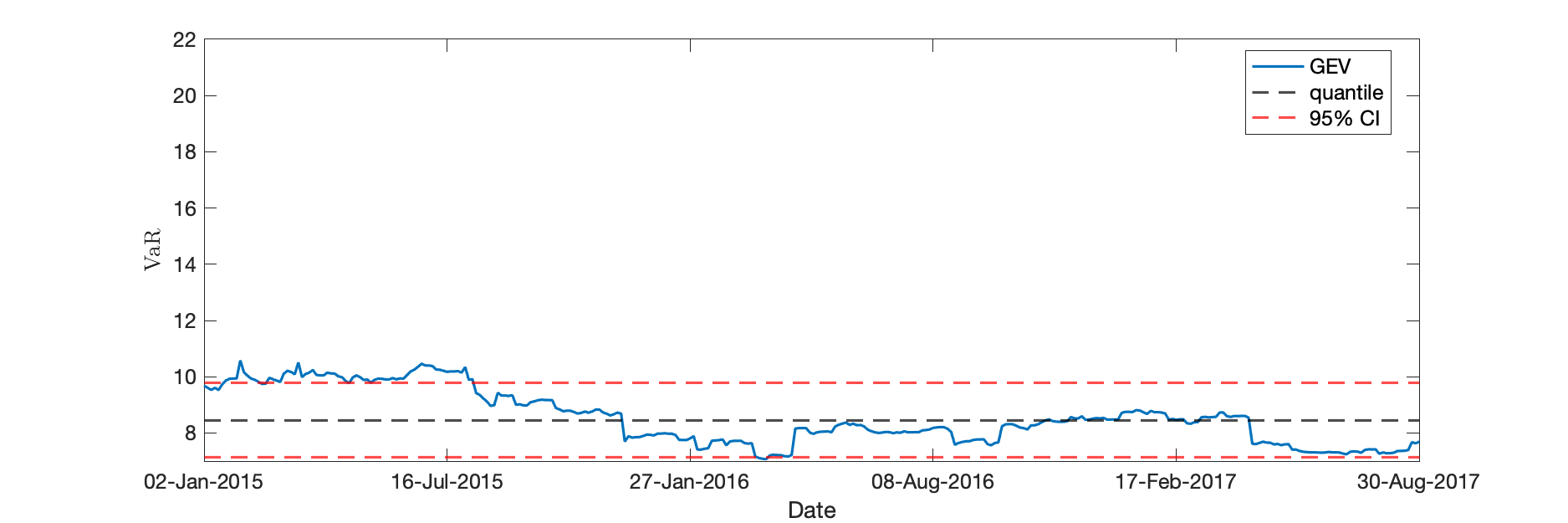}
    \caption{$VaR(s)$, CVX}
    \label{fig: cvx_q99}
  \end{subfigure}
  \caption{Dynamics of \emph{p}-value, $EVI(t)$ and $VaR(t)$ for APPL and CVX stock from 1/3/2015 to 8/31/2017. The \emph{p}-value of KS test always indicates a good fit. The $EVI(t)$ fluctuate around its mean (black). And the $VaR(t)$ are higher in 2015 than the rest of time.}
  \label{fig: dynamic_us}
\end{figure}

\subsection{Extreme value distribution comparison: Chinese vs U.S. stocks}\label{section: market compare}
\begin{figure}[!h]
  \centering
  \begin{subfigure}[b]{0.48\textwidth}
    \centering
    \includegraphics[width=\textwidth]{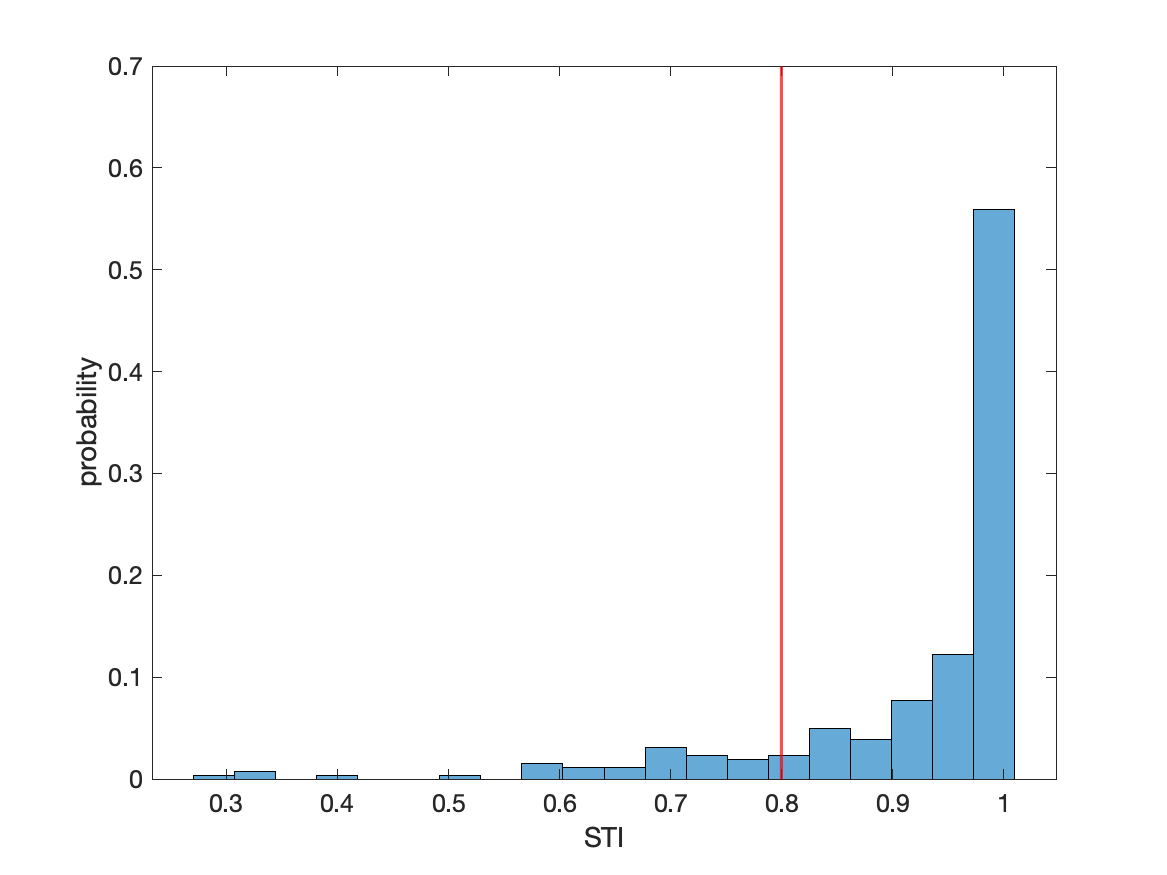}
    \caption{Chinese stock pool}
    \label{fig: chn_percent_xi}
  \end{subfigure}
  \hfill
  \begin{subfigure}[b]{0.48\textwidth}
    \centering
    \includegraphics[width=\textwidth]{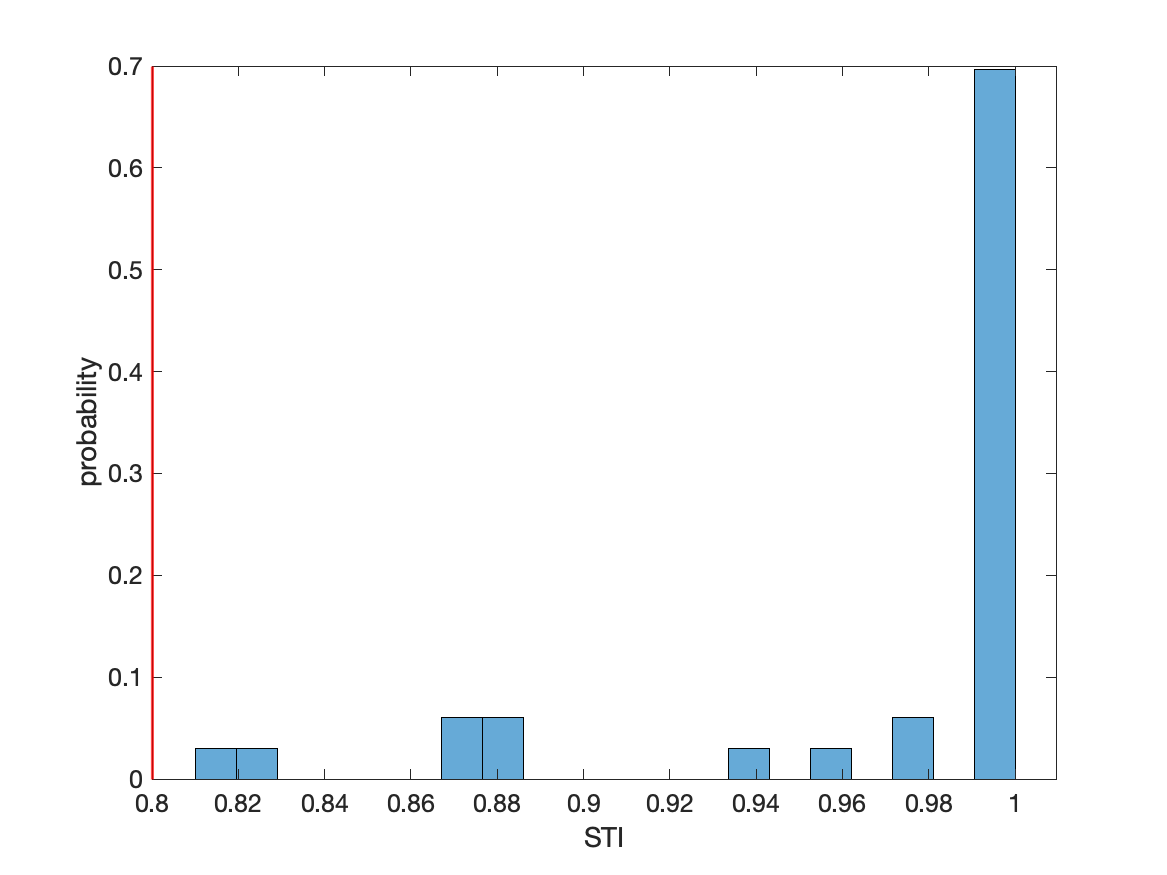}
    \caption{U.S. stock pool}
    \label{fig: us_percent_xi}
  \end{subfigure}
  \caption{Histogram for the $STI$ for 261 stocks in Chinese stock pool and 32 stocks in U.S. stock pool. In U.S. stock pool with 32 selected stocks, the $STI$ are always larger than $0.8$. These stocks have stable risk models. While in Chinese stock pool, only $85.8\%$ stocks have stable risk models.}
  \label{fig: xi_percent}
\end{figure}

For all Chinese and U.S. stocks, the \emph{p}-values of the KS test are always larger than $0.05$ and the maximum $\sup_t G_{\theta(t)}(0)$ over all stocks is $10^{-7} \ll 10^{-5}$. Both validation criteria are satisfied, giving us confidence that the GEV distribution effectively models the maxima of $|SLR_{s}|$. We found that risk models exhibit high stability for most stocks in the pool. In the U.S. stock pool, the $EVI(t)$ of all stocks remains constant over time (see Figure \ref{fig: xi_percent}), indicating stable risk models. In contrast, in the Chinese stock pool, only $85.8\%$ of stocks have stable risk models.

\begin{figure}[!h]
  \centering
  \begin{subfigure}[b]{0.48\textwidth}
    \centering
    \includegraphics[width=\textwidth]{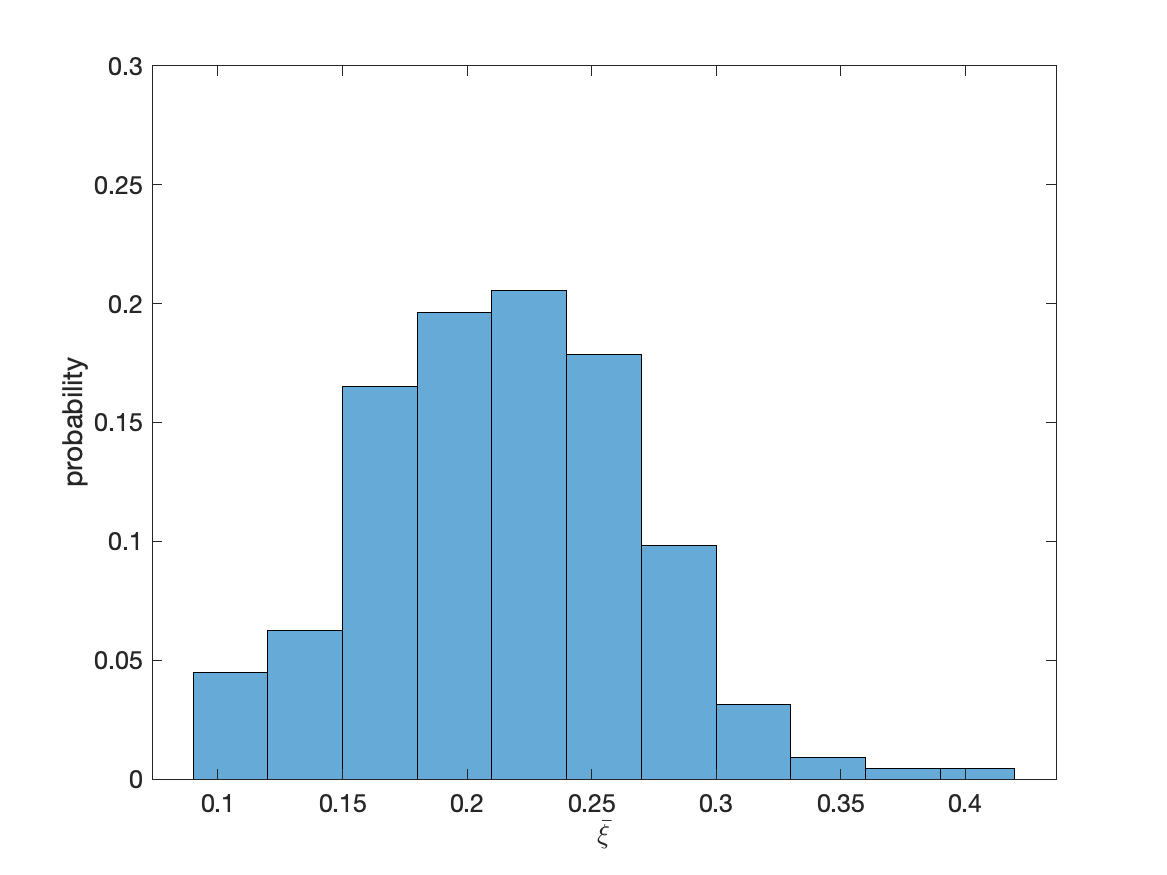}
    \caption{Chinese stock pool}
    \label{fig: chn_mean_xi}
  \end{subfigure}
  \hfill
  \begin{subfigure}[b]{0.48\textwidth}
    \centering
    \includegraphics[width=\textwidth]{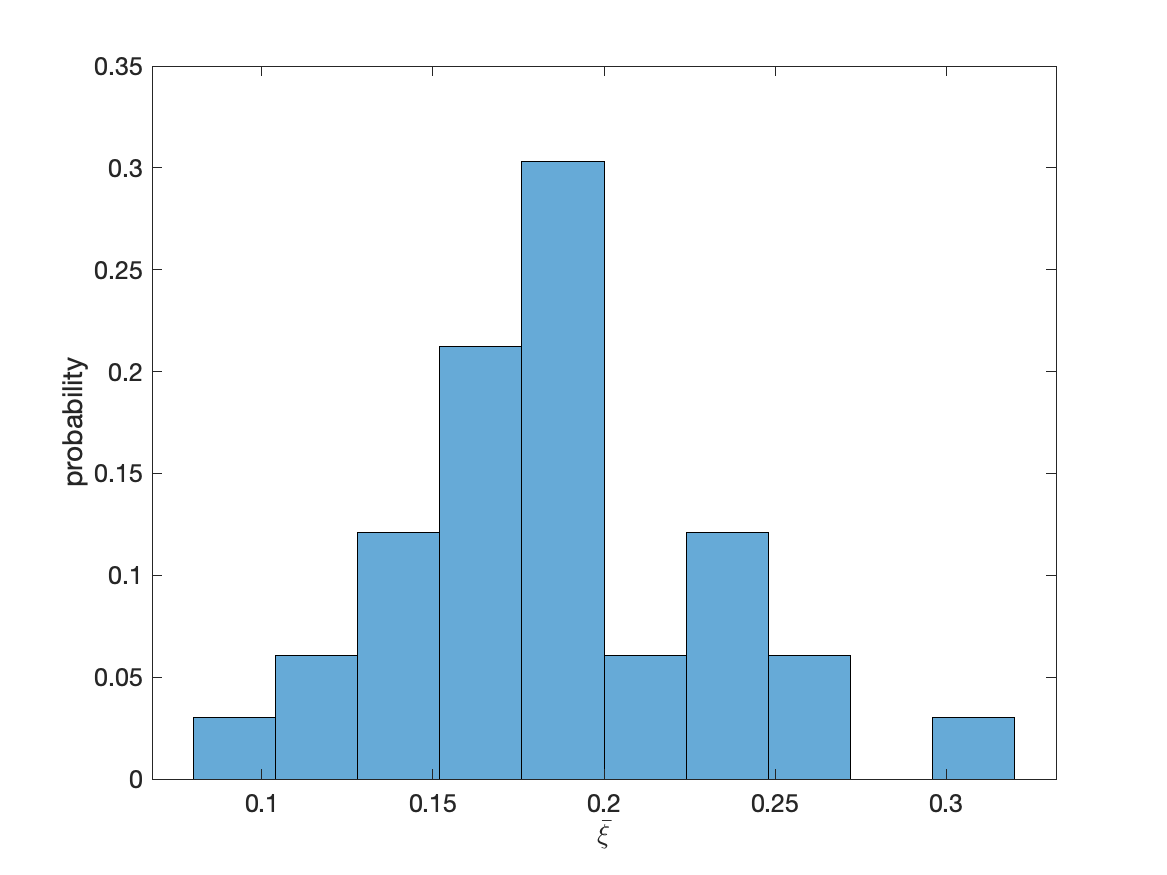}
    \caption{U.S. stock pool}
    \label{fig: us_mean_xi}
  \end{subfigure}
    \caption{Histogram of the $mEVI$ for stocks with stable risk model in Chinese and U.S. stock pool.}
    \label{fig: xi_compare}
\end{figure}

For stocks with stable risk models, we studied the estimated $mEVI$. From Figure \ref{fig: xi_compare}, we observed that the mean of $mEVI$ is $0.22 \pm 3.4 \times 10^{-3}$ for all stocks with unchanged $EVI(t)$ in the Chinese stock pool. This is significantly larger than the mean $mEVI$ ($0.17 \pm 8.4 \times 10^{-3}$) in the U.S. stock pool. Based on the tail of the GEV distribution, a larger positive $\xi$ represents a thicker upper tail, generally indicating a higher probability of extreme loss or gain. We conclude that most stocks in both the Chinese and U.S. stock pools have stable risk models, but the stocks in Chinese pool generally exhibits higher risk than the stocks in U.S. pool.

\begin{figure}[!h]
  \centering
  \begin{subfigure}[b]{\textwidth}
    \centering
    \includegraphics[width=\textwidth]{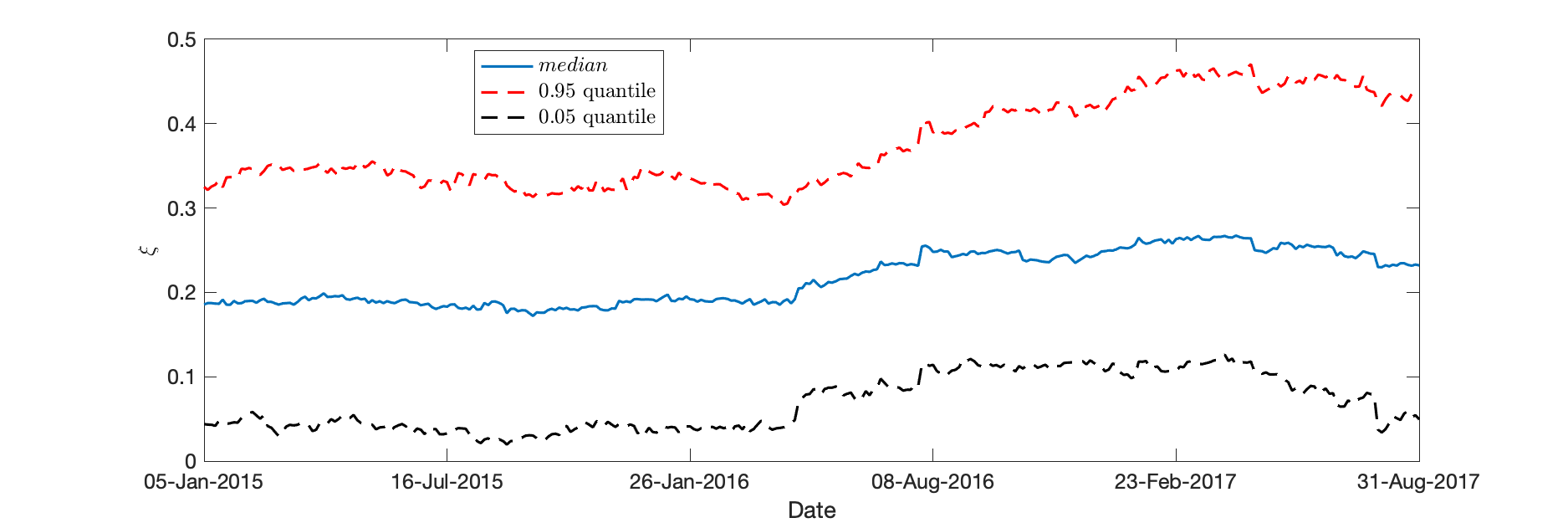}
    \caption{Chinese stock pool}
    \label{fig: chn_cross_xi}
  \end{subfigure}

  \begin{subfigure}[b]{\textwidth}
    \centering
    \includegraphics[width=\textwidth]{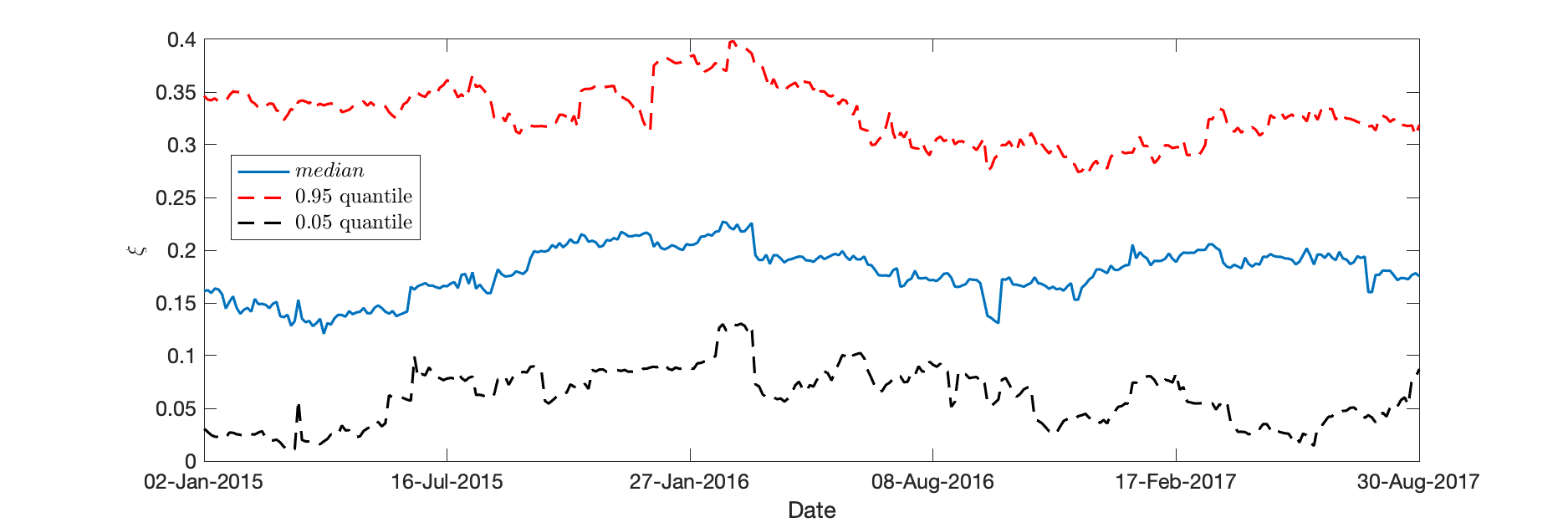}
    \caption{U.S. stock pool}
    \label{fig: us_cross_xi}
  \end{subfigure}
  \caption{Median ($mid\xi(t)$,blue), $0.05$ quantile ($low\xi(t)$, black) and $0.95$ quantile ($high\xi(t)$, red) of cross-sectional $EVI(t)$ for all stocks in Chinese and U.S. stock pool at time $t$.}
  \label{fig: market_xi}
\end{figure}

Further, we systematically monitored the $EVI(t)$ for our entire Chinese and U.S. stock pools. At time $t$, we model extreme values with the GEV distribution for each stock, resulting in a set of $EVI(t)$ values. We refer to these values as the cross-section $CRS(t)$ of $EVI(t)$ at time $t$. We then compute the range of EVI in the $CRS(t)$ and study how this range evolves over time. Specifically, we compute and display three quantiles of $CRS(t)$: the 0.05 quantile $lowEVI(t)$ (in black), the median $midEVI(t)$ (in blue), and the 0.95 quantile $highEVI(t)$ (in red). This was done separately for our Chinese and U.S. stock pools (see Figures \ref{fig: market_xi}). The $highEVI(t)$ is particularly sensitive to stocks with unstable risk models, whereas the $midEVI(t)$ is more robust to such uncertainties and generally reflects well the risk profile of the entire pool. The differences between $lowEVI(t)$, $midEVI(t)$, and $highEVI(t)$ quantify the time variability of our risk models across the pool.

We observed that the $mid{\xi}(t)$ does not change significantly in either the U.S. or Chinese stock pools. In the Chinese stock pool, $EVI(t)$ becomes slightly higher after mid-2016 compared to before. In contrast, in the U.S. market, the $mid{\xi}(t)$ increases before the beginning of 2016 and then decreases. These changes, however, are not dramatic.

From January 2014 to August 2017, we concluded that most stocks in both the Chinese and U.S. stock pools have stable risk models. Risks are generally higher in the Chinese stock pool, which also suggests a potential for larger gains. 

\subsection{VaR and a simple portfolio optimization strategy}
\begin{figure}[!h]
  \centering
  \begin{subfigure}[b]{0.48\textwidth}
    \centering
    \includegraphics[width=\textwidth]{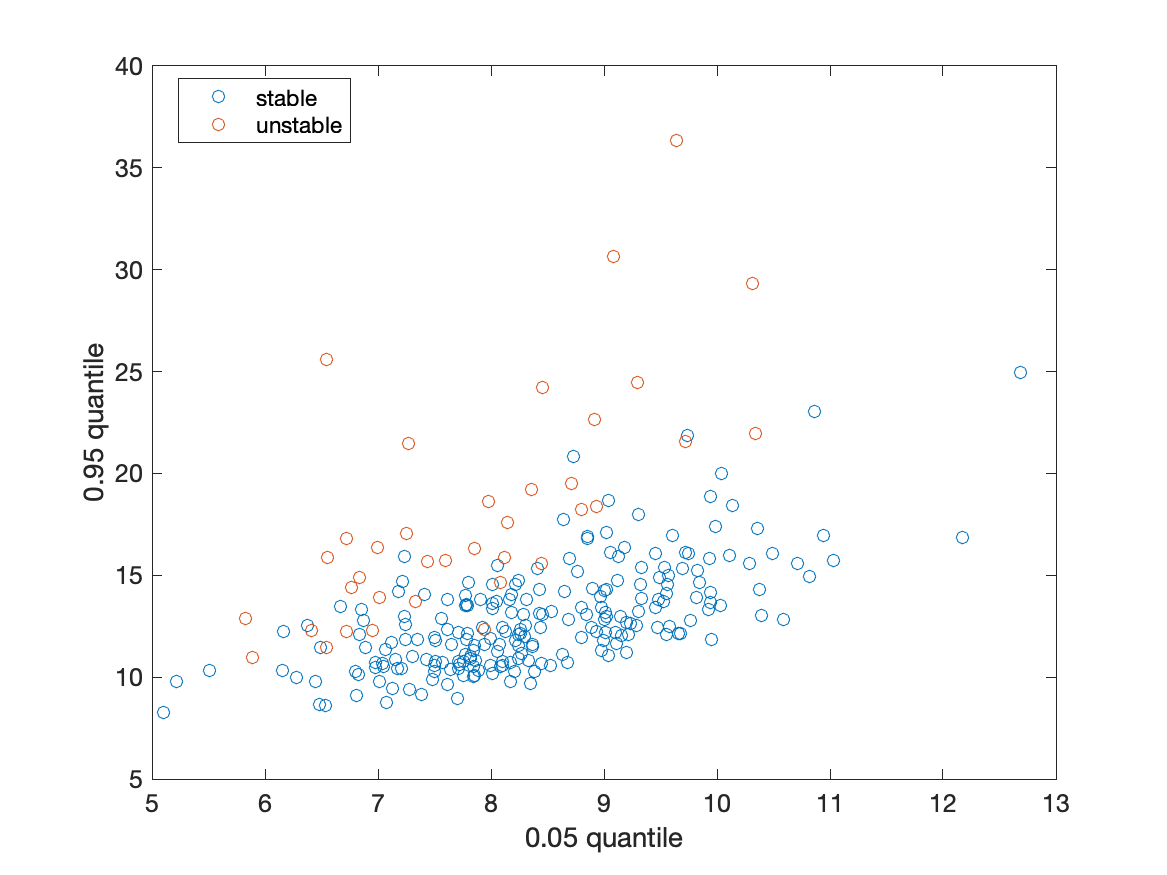}
    \caption{Chinese stock pool}
    \label{fig: chn_xi}
  \end{subfigure}
  \hfill
  \begin{subfigure}[b]{0.48\textwidth}
    \centering
    \includegraphics[width=\textwidth]{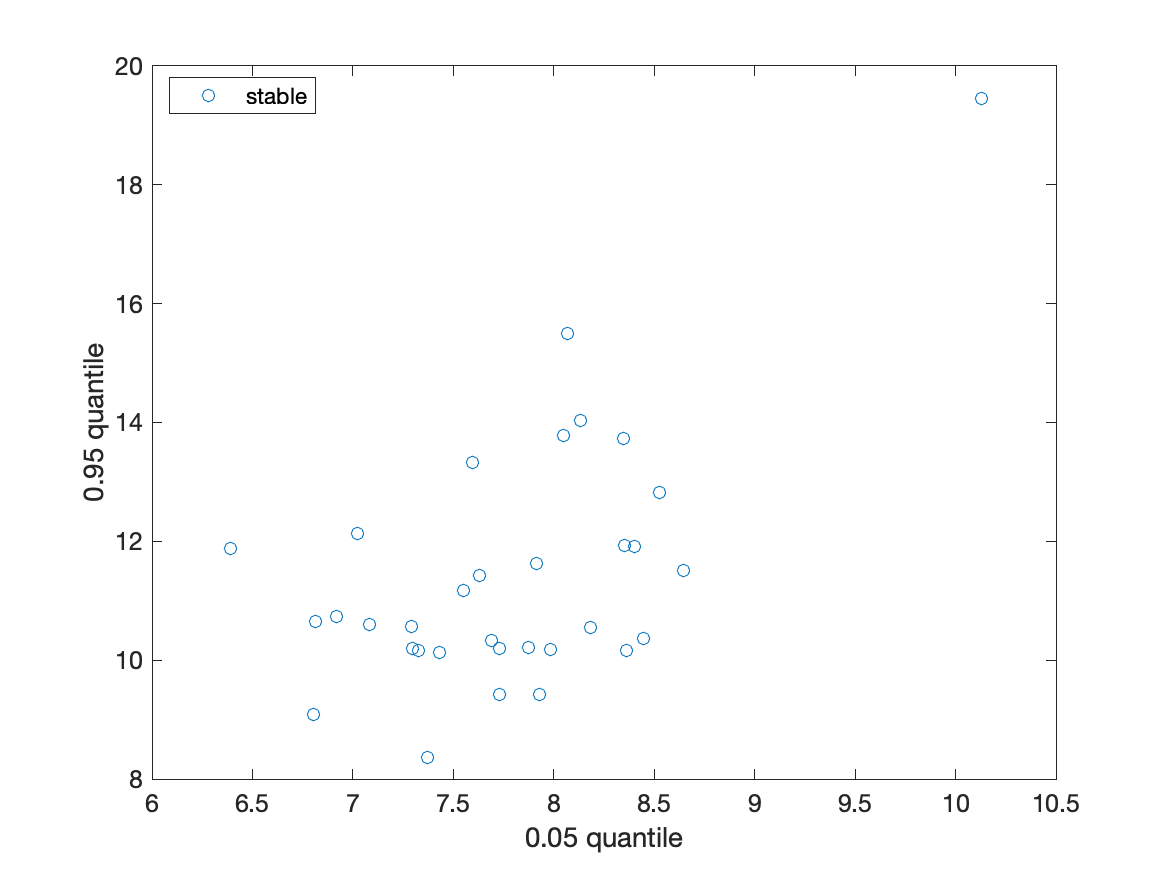}
    \caption{U.S. stock pool}
    \label{fig: us_xi}
  \end{subfigure}
    \caption{Scatter plot represents the range of $VaR(t)$ for all stocks with both stable risk models (blue) and unstable risk models (orange) in Chinese and U.S pool. The horizontal and vertical axis are $0.05$ qauntile and $0.95$ quantile of $VaR(t)$.}
    \label{fig: var range}
\end{figure}

For individual stocks with stable risk models in the Chinese stock pool, the $0.05$ quantile of $VaR(t)$ ranges from $5.1$ to $12.7$, and the $0.95$ quantile ranges from $8.2$ to $24.9$ (see Figure \ref{fig: var range}). The average difference between the $0.05$ and $0.95$ quantiles is approximately $4.5$. Similarly, for individual stocks in the U.S. stock pool, the $0.05$ and $0.95$ quantiles of $VaR(t)$ have the same range as in the Chinese stock pool. However, for stocks with unstable risk models, which are found in the Chinese stock pool, the $VaR(t)$ has a wider range. The $0.05$ quantile of $VaR(t)$ ranges from $5.8$ to $10.4$, and the $0.95$ quantile ranges from $12.8$ to $36.4$. The average difference between the $0.05$ and $0.95$ quantiles is around $10.3$.

Similar to the systematic analysis of cross-sectional $EVI$ in section \ref{section: market compare}. We also define $lowVaR(t)$, $midVaR(t)$, and $highVaR(t)$ in the cross-section at time $t$.Changes in $VaR(t)$ are more pronounced. Before analyzing these changes in $VaR(t)$ for the two pools, it is important to provide some background on the period under study. In 2014, the Chinese market surged to a peak, followed by a crash in mid-2015 due to slow GDP growth in China. This downturn had a global impact and the market did not recover until mid-2016. The stock market flourished again in 2017. The U.S. stock market experienced similar changes, influenced by the events in the Chinese market.

\begin{figure}[!h]
  \centering
  \begin{subfigure}[b]{\textwidth}
    \centering
    \includegraphics[width=\textwidth]{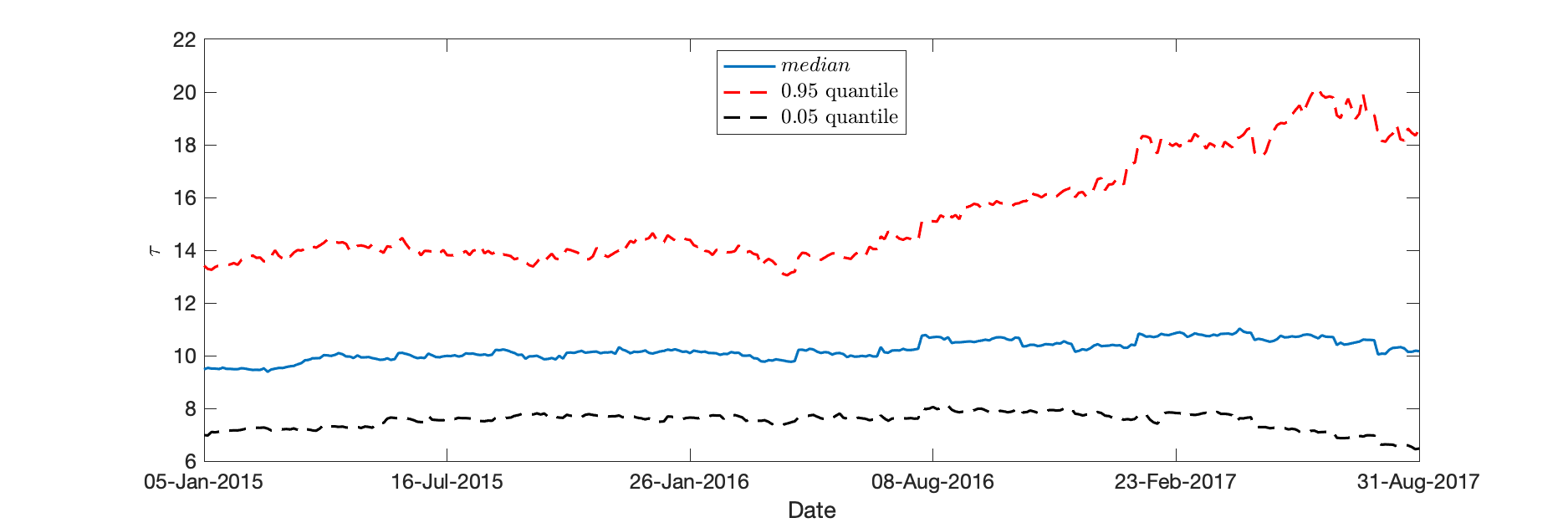}
    \caption{Chinese stock pool}
    \label{fig: chn_q99}
  \end{subfigure}

  \begin{subfigure}[b]{\textwidth}
    \centering
    \includegraphics[width=\textwidth]{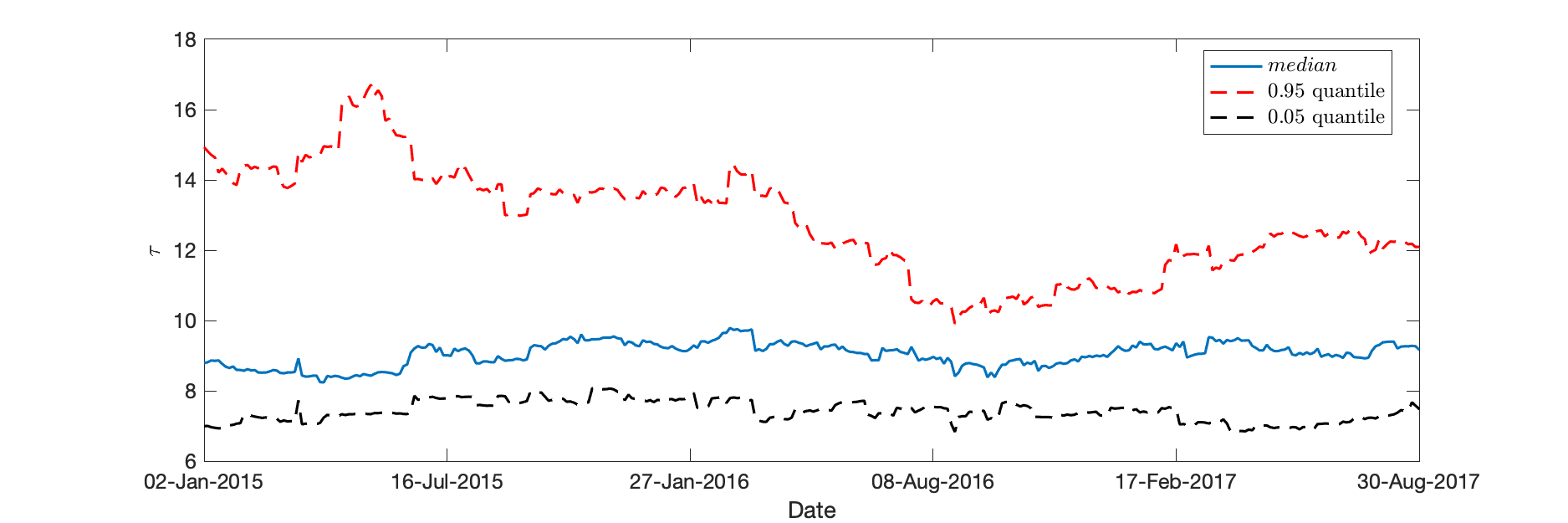}
    \caption{U.S. stock pool}
    \label{fig: us_q99}
  \end{subfigure}
  \caption{Median (blue), $0.05$ quantile (black) and $0.95$ quantile (red) of $VaR(t)$ for all stocks in Chinese and U.S. stock pool at time $t$.}
  \label{fig: market_var}
\end{figure}

Though the macro change is dramatic, it is interesting to note that the $midVaR(t)$ in both pools does not change over time, as shown in Figure \ref{fig: market_var}. However, the $highVaR(t)$ changes significantly. In the Chinese stock pool, the $highVaR(t)$ remains constant until mid-2016. During this period, the risk in both pools is almost the same. The U.S. stock pool, however, appears more sensitive to market crashes. Although the $highVaR(t)$ remains constant most of the time, there is a spike in mid-2015 that is not observed in the Chinese stock pool. Based on this observation, we consider the Chinese stock pool to be more robust to crashes. Another difference emerges from the beginning of 2016. The $highVaR(s)$ continuously increases in the Chinese stock pool, paired with a growing discrepancy in risk models, while this discrepancy decreases in the U.S. stock pool. We found that the U.S. stock pool generally exhibits lower risk after mid-2016. Conversely, risk increases in the Chinese stock pool as the economy flourishes. 

VaR is one of the most important features to consider while constructing a portfolio (\cite{hidayana2023}). We display a simple weights calculation method to optimize a portfolio based on these cross-sectional VaR and compare precisely the advantage of our GEV-VaR to the Normal-VaR. The weights $w_i(t)$ of stocks $i$ in the pool is calculated by:
\begin{equation*}
    w_i(t) = \frac{\exp\left(-VaR_i(t)\right)}{\sum_i \exp\left(-VaR_i(t)\right)}
\end{equation*}

The example elaborated below consider both the portfolio of 261 Chinese stocks and 32 U.S. stocks. Note that we adjust the weights of stocks in portfolio every $22$ days, which is equivalent to 1 trading month. We compute the weights on the day 22 based on the current $VaR(t)$ and update the portfolio on day 1 of every period.

\begin{figure}[!h]
    \centering
    \includegraphics[width=0.7\textwidth]{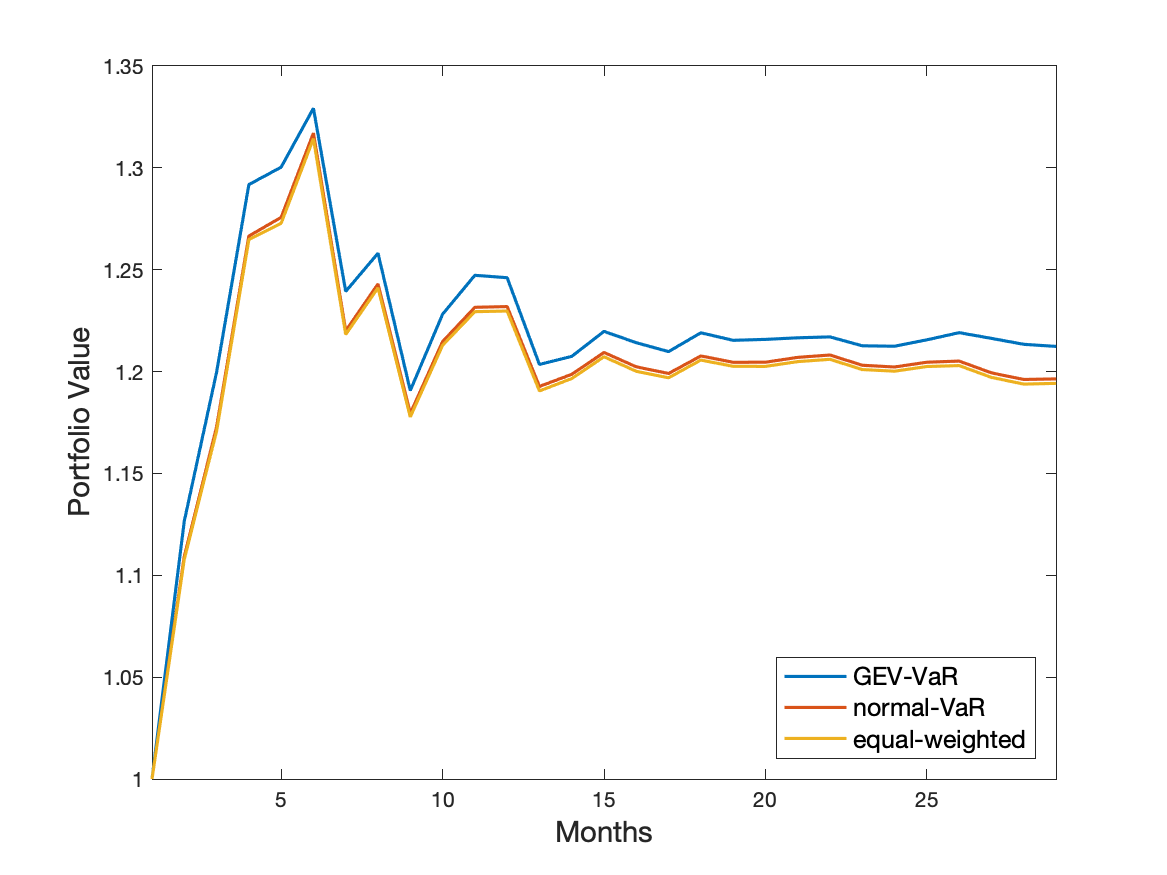}
    \caption{Portfolio values from 2015/01/05 to 2017/08/31 for GEV-VaR portfolio (blue), Normal-VaR portfolio (red), and equal-weighted portfolio (orange) consist of 261 Chinese stocks (Figure \ref{fig: CHN portfolio}). Weights in GEV-VaR portfolio are updated monthly based on our GEV-VaR, weights in Normal-VaR portfolio are updated monthly based on the Normal-VaR. Monthly position reduction plan is adopted here.}
    \label{fig: CHN portfolio}
\end{figure}

We first constructed three portfolios for all stocks in Chinese stock pool. GEV-VaR portfolio(blue) has weights updated monthly based on GEV-VaR. Normal-VaR portfolio (red) has weights updated monthly based on Normal-VaR. While equal-weighted portfolio (orange) allocates equal weights to individual stocks in the portfolio all the time. Besides, we reduce positions month by month that we only hold $1/{number\ of\ month}$ of total position to prevent the loss in potential bear market based on the fact that the Chinese market has been bull for almost one year. We implemented a backtest, and the results in Figure \ref{fig: CHN portfolio} show that the GEV-VaR based portfolio consistently achieves higher portfolio value and slightly higher returns than both the Normal-VaR portfolio and the equal-weighted portfolio, with an increase of $2.1\%$ by 2017/08/31. Specifically, the GEV-VaR portfolio attained the highest return of $21.5\%$ by the same date. It is $27.1\%$ higher than the Shenzhen Component Index.

\begin{figure}[!h]
    \centering
    \includegraphics[width=0.7\textwidth]{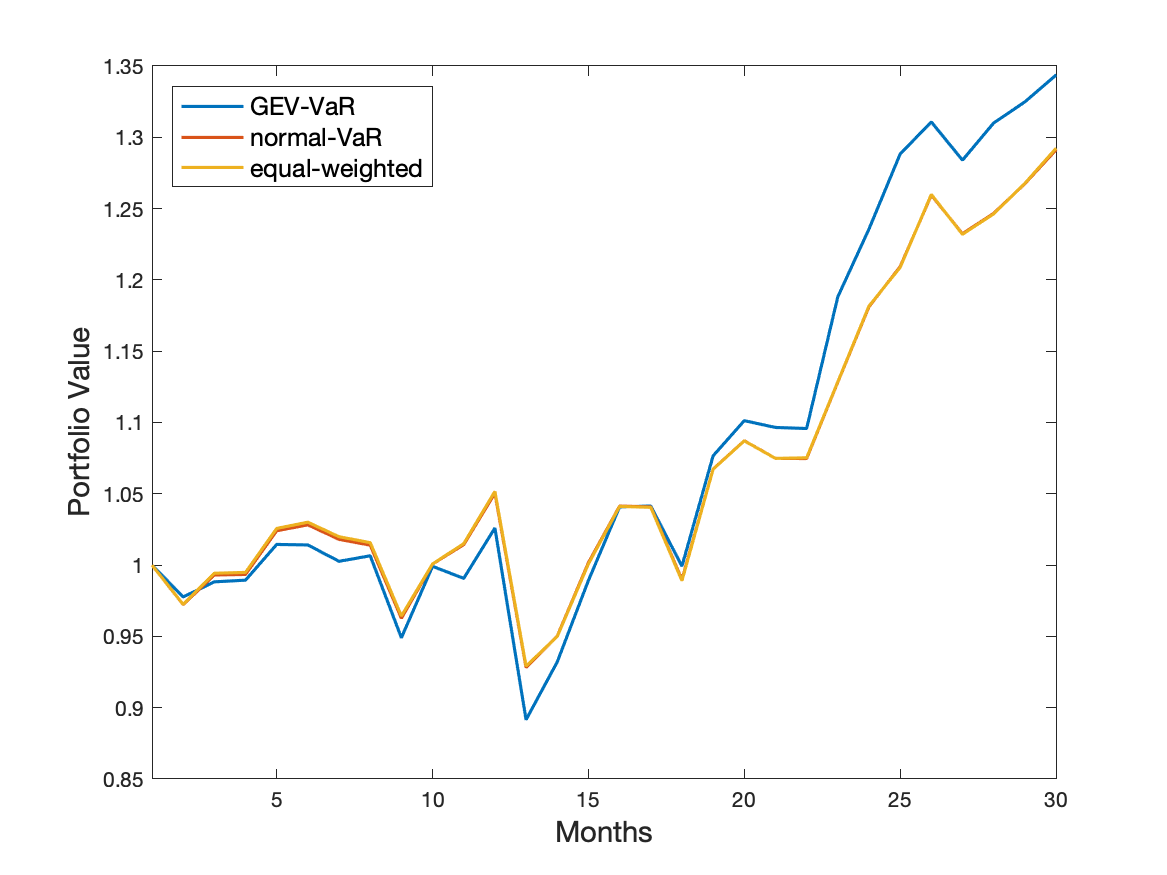}
    \caption{Portfolio values from 2015/01/05 to 2017/08/31 for GEV-VaR portfolio (blue), Normal-VaR portfolio (red), and equal-weighted portfolio (orange) consist of 32 U.S. stocks (Figure \ref{fig: us portfolio}). Weights in GEV-VaR portfolio are updated monthly based on our GEV-VaR, weights in Normal-VaR portfolio are updated monthly based on the Normal-VaR.}
    \label{fig: us portfolio}
\end{figure}

Throughout the evaluation period, the Normal-VaR portfolio's performance closely mirrored that of the equal-weighted portfolio. The key distinction between the GEV-VaR and Normal-VaR portfolios lies in their treatment of stock weights. The Normal-VaR approach results in minimal differences among stock weights, failing to capture the tail behavior accurately. In contrast, our GEV-VaR method, by identifying higher Value at Risk (VaR) values, assigns lower weights to riskier stocks, thus effectively managing risk. This differentiation ensures that low-risk stocks maintain a stable risk profile within the portfolio. Consequently, as the market appreciates, our portfolio value increases. This pattern is also evident in the U.S. portfolio results shown in Figure \ref{fig: us portfolio}, where the values of the Normal-VaR and equal-weighted portfolios almost overlap, underscoring the uniform weighting approach's limitations.

We adopted the same strategy to construct the U.S. portfolio during the same period, except for the monthly position reduction plan, as the surge in the U.S. market in 2014 was not dramatic. Still, we found our GEV-VaR driven portfolio (GEV-VaR portfolio) achieved the largest return ($34.3\%$) by 2017/08/31, which is $5.1\%$ higher than the normal VaR-based portfolio (Normal-VaR portfolio) and the equal-weighted portfolio (equal-weighted portfolio). The gains of GEV-VaR based portfolios exceeded the Dow Jones Industrial Average by $13.0\%$.

\section{{ Validation of our GEV approach on  simulated  Heston joint SDEs}}\label{section: simulation}

{Stochastic volatility models for stock log-prices have been extensively studied by \cite{bauwens2004, aitsahalia2009, cont2011, chen2017}, leading to the commonly used SDE model:}
\begin{equation*}
d\log P_s = \mu_s ds + \sqrt{V_s} dB_s
\end{equation*}
{where $P_s$ is the asset price at time $s$, $\mu_s$ is a deterministic trend, $V_s$ is the squared spot volatility, and $B_s$ is a standard Brownian motion.}

{For the classical Heston model, the squared volatility $V_s$ is driven by another SDE:}
\begin{equation*}
dV_s = \kappa (\theta - V_s) ds + \gamma \sqrt{V_s} dW_s
\end{equation*}

{Here, $B_s$ and $W_s$ are two standard Brownian motions with constant instantaneous correlation $\rho$, so that $E[dB_s dW_s] = \rho ds$. The positive parameters $\kappa, \theta, \gamma$ must satisfy the well-known Feller condition $\kappa\theta/\gamma^2 > 1/2$, which ensures the almost sure positivity of $V_s$ for all $t > 0$.}

{In a high-frequency setting, the impact of the drift term is negligible (see \cite{barndorff2004}). Thus, we simply assume $\mu_s = 0$. Then $LR_s = d\log P_s$ and $V_s$ verify the Heston joint SDEs:}
\begin{align}\label{eq: heston}
&LR_s = d\log P_s = \sqrt{V_s} dB_s \\
&dV_s = \kappa (\theta - V_s) ds + \gamma \sqrt{V_s} dW_s
\end{align}


\subsection{Canonical volatility process}

{By applying a linear scaling factor $A = \kappa / \gamma^2$ to the volatility values $V_t \to A\,V_t$ and a time rescaling factor $t \to \kappa t$, \cite{azencott2015} introduced a \emph{canonical form} for the Heston volatility SDE, namely:}
\begin{equation}\label{eq: canonical SDE}
dV_s = (z - V_s) ds + \sqrt{V_s} dH_s
\end{equation}
{where the single parameter $z = \kappa\theta / \gamma^2$ and $H_s$ is a new standard Brownian motion. Since the initial Heston parameters $\kappa, \theta, \gamma$ satisfy the Feller condition $\kappa\theta / \gamma^2 > 1/2$, the single parameter $z$ of a canonical SDE always satisfies $z > 1/2$. The canonical SDE is of course still a Heston SDE parameterized by $\kappa = 1$, $\theta = z$, and $\gamma = 1$. The canonical return process $LR_s$ can be recovered from the canonical process $V_s$ by the SDE}
\begin{equation}
    LR_s = \sqrt{V_s} dB_s
\end{equation}

{The canonical Heston SDEs are quite convenient to study the extreme value distribution of the return process since it has only a single parameter. The extreme value results for the canonical Heston SDE can be easily transformed into results for the generic Heston SDEs because linear changes in scale and linear time subsampling will simply modify the location and scale parameters in the fitted extreme value distributions. Therefore, for our intensive Heston SDE simulations, we will use the canonical form of the Heston volatility SDE. More technical details will be expanded in Section \ref{section: BM}.}

\subsection{Time discretization}
To simulate a canonical Heston SDE driving the log-returns $LR_s$ and their volatility $V_s$ over a time interval $[0, T]$, we discretize time with equal time steps of extremely small size $\epsilon$.

We conduct extensive simulations of the pair of canonical Heston SDEs for various values of the parameter $z > 0$. The traditional Euler scheme is not sufficiently accurate for these simulations, particularly when $z$ is close to $1/2$. Instead, we employ the implicit Milstein scheme (\cite{doss1977, kahl2006}) to simulate the volatility. This approach ensures that the simulated volatility remains positive when $z > 1/2$ and offers first-order strong convergence, with the expected absolute difference between simulated terminal values and reference being $O(\epsilon)$.

For each specific Heston parameter $z$, we simulate long trajectories of the corresponding Heston SDEs, with duration $T = 896$ and $\epsilon = 1/14400$. 

\subsection{Standardized log-return}

{Fix a time scale $\Delta \gg \epsilon$, and define the corresponding discretized log-returns $LR_{h}$ observed at the discrete instants $h= h_j = j \Delta$, by 
\begin{equation}
     LR_h = \log P_{h+\Delta}-\log P_h 
\end{equation}. 
For instance, if our Heston SDEs had been fitted to real intraday data, we could take $\epsilon$ as 1 second for simulations, and the time scale $\Delta$ of successive observations could be 30 seconds, 1 minute, or 5 minutes.}\\

{At time scale $\Delta$ , for each observation time    $h=h_j =j \Delta$, define the standardized log-return $SLR_h$  by }
\begin{equation}
 SLR_h = \frac{LR(h) - E(LR(h)}{std(LR_h)}
\end{equation}

For the Heston SDEs \eqref{eq: heston}, one has $E\left(LR_h\right) = 0$, and for small time scale $\Delta$, the standard deviation $std(LR_h)$ can be approximated by $\sqrt{V_h \Delta}$. But for large time scale $\Delta$:
\begin{equation*}
std(LR_h) = \sqrt{\int_{h}^{h+\Delta}V_u du}
\end{equation*}

\subsection{Numerical analysis of extreme values  for  simulated standardized log-returns}

{For each pre-selected canonical Heston parameter $z$ and each pre-selected time scale $\Delta = 1/240, 1/48, 1/24, 1, 2$, we compute the observed $SLR_{h}$. We conjecture that each $|SLR_h|$ has an approximately  standard normal distribution, and we have validated this conjecture empirically (see Table \ref{table: dist L}).} Our statistical test of normality is here the classical  one-sample KS test. As is well known, normal distributions belong to the max-attraction domain of the Gumbel distribution, i.e., the corresponding GEV has shape parameter $\xi = 0$ (\cite{beirlant2005estimation}).

\begin{table}[!h]
\centering
\resizebox{0.6\textwidth}{!}{%
\begin{tabular}{|l|r|r|r|}
\hline
\multicolumn{1}{|r|}{} & \multicolumn{1}{c|}{range of \emph{p}-value} & \multicolumn{1}{c|}{} & \multicolumn{1}{c|}{range of \emph{p}-value} \\ \hline
$z=0.55$               & [0.41, 0.99]                                 & $z=4$                 & [0.38, 0.98]                                 \\ \hline
$z=1$                  & [0.06, 0.36]                                 & $z=5$                 & [0.18, 0.83]                                 \\ \hline
$z=1.5$                & [0.07, 0.56]                                 & $z=6$                 & [0.16, 0.75]                                 \\ \hline
$z=3$                  & [0.37, 0.71]                                 & $z=7$                 & [0.14, 0.75]                                 \\ \hline
\end{tabular}%
}
\caption{Range of \emph{p}-values for normality test of $SLR_h$ (by one-sample KS test), when $\Delta = 1/240, 1/48, 1/24, 1, 2$ and various $z$. These \emph{p}-values are always larger than $0.05$, so we conjecture that the $SLR_h$ approximately always has a normal distribution.}
\label{table: dist L} 
\end{table}

{For each fixed $\Delta$, we partition the simulated long trajectory into 448 blocks of equal duration 2, indexed by $t=1,...,448$, and we extract the maximum of $SLR_{h}$ in every block. At each time $t$, we fit a GEV model to the maxima of $SLR_{h}$ extracted from blocks $t,t-1,...,t-122$. The estimated GEV parameters are denoted $EVI_{z}(t)$, $\hat{\mu}_{z}(t)$, and $\hat{\sigma}_{z}(t)$ for all $z$ and $t$, and of course depend on the parameter $\Delta$. For each fixed $\Delta$ as above, the preceding estimation procedure is repeated for 600 simulations of each one of our eight canonical Heston models indexed by $z= 0.55,1,1.5, 3,4,5,6,7$. Estimation results are given in Table \ref{table: heston_GEV}.}

To validate our extreme value analysis framework, we address three specific questions:
\begin{enumerate}
\item Can the GEV distribution correctly model $\max_h\left(|SLR_h|\right)$ for each canonical Heston parameter $z$ and each $\Delta$?
\item For fixed $z$ and $\Delta$, does the shape parameter $EVI_{z}(t)$ change over time?
\item If $EVI_{z}(t)$ is statistically constant over time, what are the ranges of values for $mEVI_{z}$ and $VaR_{z}$?
\end{enumerate}

{To answer the first question for each fixed time scale $\Delta$, as outlined in Section \ref{section: framework}, we compute at each time $t$ the \emph{p}-value $pval(t)$ of our goodness of fit test, and the GEV model positivity index MPI(t). For all our simulated 19,200 long trajectories of Heston models, $pval(t) > 0.05$ in 99.99\% of cases, and for all these 19,200 trajectories, we always have $MPI(t) < 10^{-13}$. This confirms the systematic validity of the GEV model for the maxima of $|SLR_h|$ across all simulated trajectories of the Heston SDEs and for all values of $\Delta$.}

{To address the second question, we computed the stability index (STI) of $EVI_{z}(t)$ for each simulated long trajectory of our Heston models. For fixed $\Delta$, we found that 95\% of the 19,200 $EVI_{z}(t)$ trajectories have an STI greater than $80\%$, indicating that $EVI_{z}(t)$ remains statistically constant. This was to be expected since the absolute standardized returns are stochastically stationary over time.}

{To answer the third question for each fixed $\Delta$, we computed the mean values $mEVI_{z}$, $\bar{\mu}_{z}$, and $\bar{\sigma}_{z}$. We found that when $\Delta$ is fixed, these three mean values remain statistically unchanged when the canonical Heston parameter $z$ varies, as tested by a t-test. However, when $\Delta$ varies, these three mean values change significantly, in part because $\Delta$ determines the number of observations $m=1/\Delta$ in each block. This aligns with the results of Table \ref{table: dist L}, which show that the distributions of standardized returns are approximately normal. Indeed, the $mEVI_{z}$ converge to zero slowly as $\Delta$ decreases, consistent with theoretical expectations (\cite{bibinger2021}).}

{As indicated by our Table \ref{table: heston_GEV}, we found that the shape parameters $EVI_{z}(t)$ fitted to our simulated Heston SDEs increase from $-0.11$ to $-0.08$ as $\Delta$ decreases from $1/24$ to $1/240$. These changes are significant under a t-test. The standard error of estimation on $EVI_{z}(t)$ remains of order $0.001$ throughout. As $\Delta$ decreases, $\bar{\mu}_{z}$ increases from $2.31$ to $3.08$, and $\bar{\sigma}_{z}$ decreases from $0.37$ to $0.30$. The $VaR$ is computed as the $0.99$ quantile of $G_{\bar{\theta}}$, and increases as $\Delta$ shrinks to 0.}

{We have also fit  Generalized Pareto (GP) distributions to one of our simulated Heston models, with satisfactory goodness of fit. Across our 600 long Heston trajectories, the VaR estimates derived from fitted GP distributions were compared to the VaR estimates derived from fitted GEV distributions. These two sets of VaR estimates had quite close empirical means, but we noted that our GEV-based VaR estimates had smaller empirical variance than the GP-based VaR estimates.}

\begin{table}[!h]
\centering
\resizebox{0.4\columnwidth}{!}{%
\begin{tabular}{|l|r|r|r|r|}
\hline
                 & \multicolumn{1}{c|}{$mEVI$} & \multicolumn{1}{c|}{$\bar\mu$} & \multicolumn{1}{c|}{$\bar\sigma$} & \multicolumn{1}{c|}{$VaR$} \\ \hline
$\Delta = 1/240$ & -0.08                          & 3.08                           & 0.30                              & 4.23                       \\ \hline
$\Delta = 1/48$  & -0.10                          & 2.56                           & 0.35                              & 3.85                       \\ \hline
$\Delta =1/24$   & -0.11                          & 2.31                           & 0.37                              & 3.67                       \\ \hline
\end{tabular}%
}
\caption{{Mean values $mEVI_{z}$, $\bar{\mu}{z}$, and $\bar{\sigma}{z}$. For fixed $\Delta$ and Heston parameter $z$, we first average $EVI_{z}(t)$ over time for each one of our 600 Heston trajectories. The mean of these 600 time averages is denoted $mEVI_{z}$. A similar procedure is used to compute $\bar{\mu}{z}$ and $\bar{\sigma}{z}$. Standard errors are of the order $10^{-3}$ for all these three estimates. For fixed $\Delta$, the results are the same for $z=0.55,1,1.5,3,4,5,6,7$, but they are definitely impacted by changes in the time scale $\Delta$.}}
\label{table: heston_GEV}
\end{table}

\section{A brief discussion on extreme values models applied to stock prices jumps detection}\label{section: jump}

{  Jumps in high-frequency stock prices $P_s$ have often been  modeled by adding an  extra component $X_s$ in SDE models of log-returns with stochastic volatility $V_s$, as follows } 
\begin{equation*}
d\log P_s = \sqrt{V_s} dB_s + X_s dJ_s
\end{equation*}
{ Here $X_s$ denotes the random price jump size and $dJ_s$ the increment of a discrete count process. \cite{lee2008jumps} 
adopted a Gumbel GEV distribution with associated  specific re-normalization parameters to model, in the absence of jumps,  the maximum behaviour of $|SLR_h|$ at various time scales $\Delta$. Indeed the paper had first showed that for \textbf{vanishing time scale} $\Delta\to 0$, one could model the distribution of $SLR_h$ by a standard normal distribution. This approach led to use the $0.99$ quantile of the Gumbel distribution as a threshold to detect jumps.}

\begin{figure}[!h]
  \centering
  \begin{subfigure}[b]{\textwidth}
    \centering
    \includegraphics[width=\textwidth]{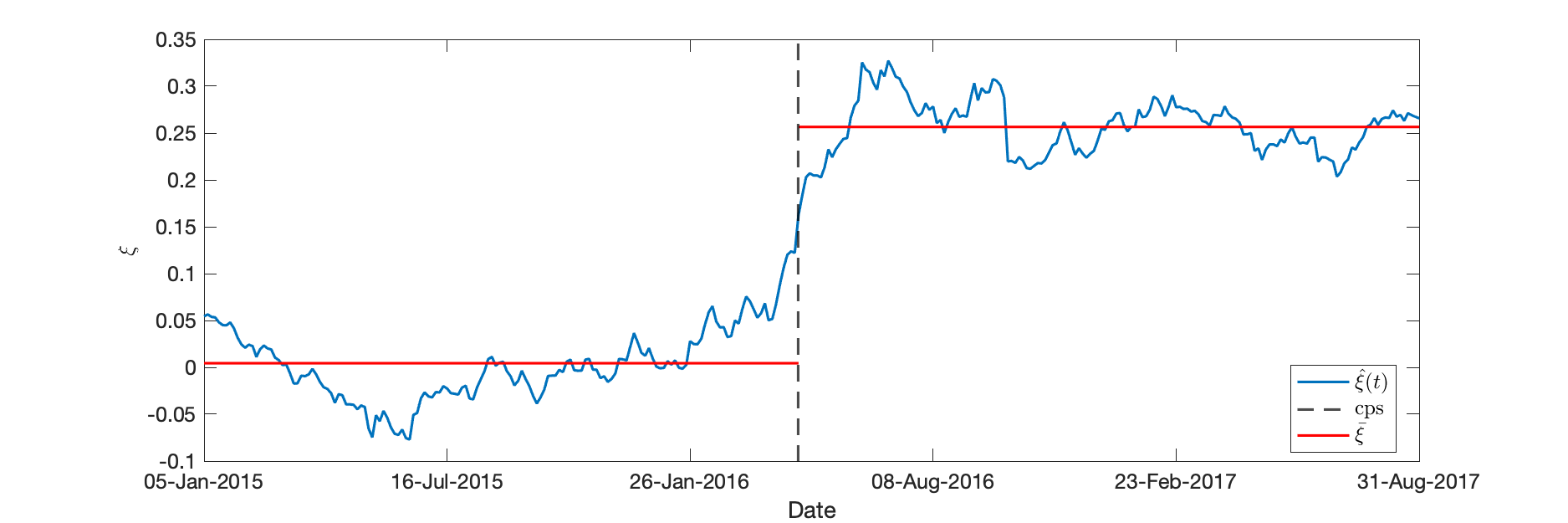}
    \caption{$EVI(t)$}
    \label{fig: cps_evi}
  \end{subfigure}

  \begin{subfigure}[b]{\textwidth}
    \centering
    \includegraphics[width=\textwidth]{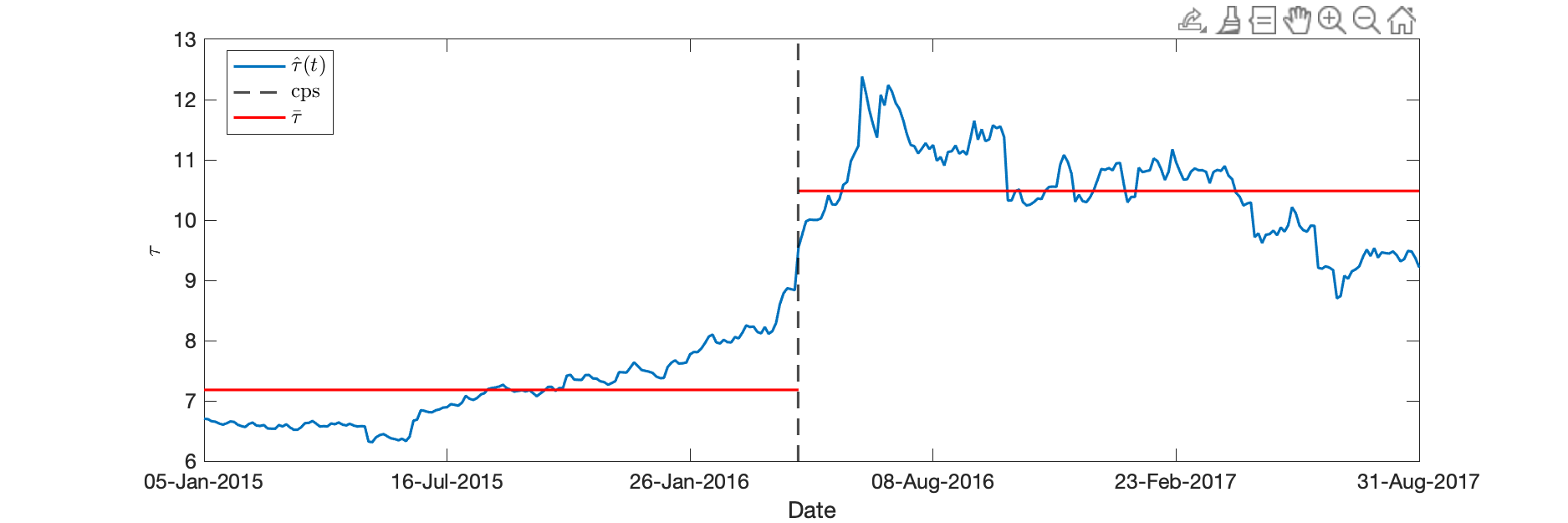}
    \caption{$0.99$ quantile}
    \label{fig: cps_var}
  \end{subfigure}
  \caption{Change point (black) detected in $EVI(t)$ (blue), paired with a change in $0.99$ quantile (blue) for stock 300193. Change points are detected by BOCD algorithm.}
  \label{fig: stock_xi_change}
\end{figure}

{However, there are several difficulties in the extensive use of the Gumbel distribution to detect jumps. Indeed in our study of extreme values for real intraday data on the Chinese and the US stock markets,  we have shown above that \textbf{even at moderately large time scales } $\Delta$, the distribution of maxima for $SLR_h$ could not be correctly fitted the by Gumbel distribution, but required the use of GEV distributions with shape parameters EVI very different from 0. Moreover we have also shown that the GEV models correctly fitted to intraday data over long moving time windows may very well slowly change over time (see Figure \ref{fig: stock_xi_change}).}

{For the diffusion processes modeled by Heston SDEs, our extensive simulations and normality tests have led us to conjecture that even when the time scale $\Delta$ becomes quite large, the $SLR_h$ still seem to have a standard normal distribution (see Section \ref{section: simulation}).}

{In view of these two issues with Gumbel distribution modeling of extreme values for log-returns, we propose to detect jumps by first fiting an adaptive GEV model to data, with parameters slowly changing over time. Our approach would be well suited to the Chinese stock market, where we have noted that our GEV risk models do change over time. To detect and/or anticipate large jumps, the large jump threshold could naturally be our estimate of the value at risk VaR , derived from a GEV risk model dynamically adjusted to a long moving time window.}

\begin{figure}[!h]
  \centering
  \begin{subfigure}[b]{0.47\textwidth}
    \centering
    \includegraphics[width=\textwidth]{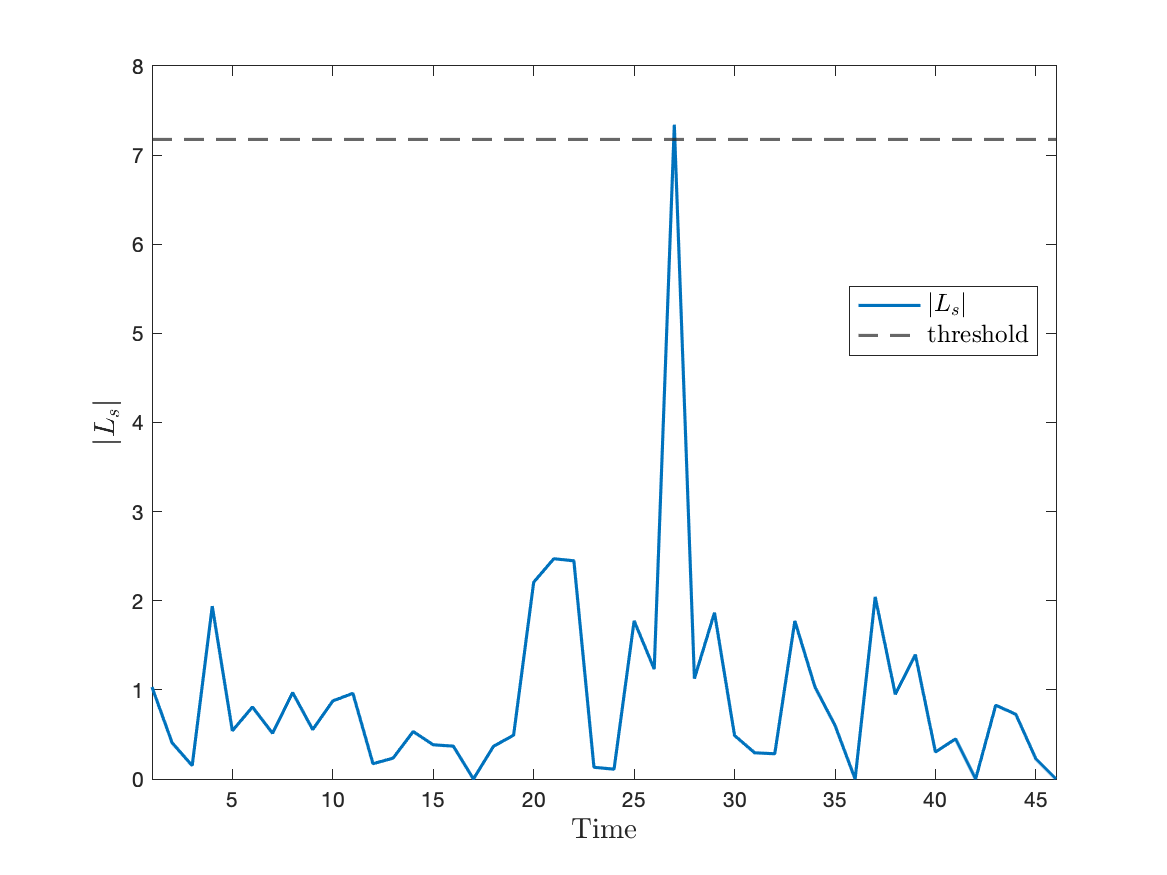}
    \caption{$SLR_h$}
    \label{fig: J_return}
  \end{subfigure}
\hfill
  \begin{subfigure}[b]{0.47\textwidth}
    \centering
    \includegraphics[width=\textwidth]{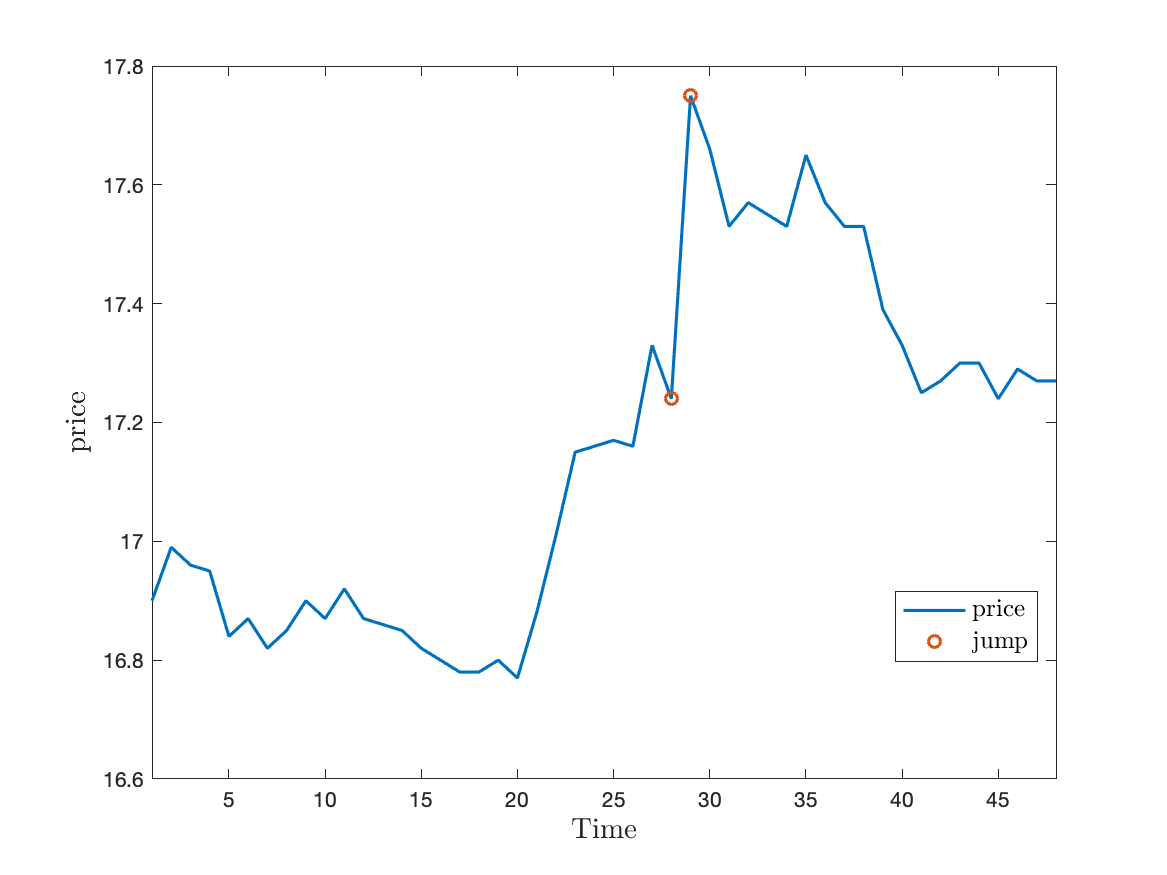}
    \caption{Price}
    \label{fig: J_price}
  \end{subfigure}
  \caption{large jump detected in 10-min frequency stock price in stock 300193 on 2015/11/30.}
  \label{fig: large jump}
\end{figure}

{Let us outline one example of this approach, for the Shenzhen Jasic Technology stock. After our dynamic fitting of GEV models to extreme values of $SLR_h$, we apply the well known BOCD (Bayesian Online Changepoint detection) algorithm to detect a change in our estimated GEV parameters on 2016/04/21, paired with a change in the estimated VaR (see Figure \ref{fig: stock_xi_change}). The shape parameter $EVI(t)$ fluctuated around or slightly below $0$ before 2016/04/21, and then started to increase up  to $0.27$. Our estimated VaR was $7.2$ before 2016/04/21, and increased to $10.3$ after that date. Since VaR was our dynamic jump detection theshold, only one large jump in stock price  was detected (on 2015/11/30) over the whole period (see Figure \ref{fig: large jump}). }

\begin{acks}[Acknowledgments]
The authors would like to thank the anonymous referees, an Associate
Editor and the Editor for their constructive comments that improved the
quality of this paper.
\end{acks}

\begin{funding}
The first author was supported by NSF Grant DMS-??-??????.

The second author was supported in part by NIH Grant ???????????.
\end{funding}

\end{document}